\documentclass[twocolumn]{aastex701}
\usepackage{multirow}
\usepackage{amsmath}
\usepackage{comment}
\usepackage{threeparttable}

\newcommand{\co}{\mbox{$^{12}$CO (2--1)}}
\newcommand{\tco}{\mbox{$^{13}$CO (2--1)}}
\newcommand{\cho}{\mbox{C$^{18}$O (2--1)}}
\newcommand{\sio}{\mbox{SiO (5--4)}}
\newcommand{\metanol}{\mbox{CH$_3$OH (4$_2$--3$_1$)}}
\newcommand{\so}{\mbox{SO (6$_5$--5$_4$)}}
\newcommand{\dcn}{\mbox{DCN (3--2)}}
\newcommand{\Cyclopropenyo}{\mbox{c-C$_3$H$_2$ (6$_{0,6}$--5$_{1,5}$)}}
\newcommand{\Cyclopropenyt}{\mbox{c-C$_3$H$_2$ (5$_{1,4}$--4$_{2,3}$)}}
\newcommand{\Cyclopropenytt}{\mbox{c-C$_3$H$_2$ (5$_{2,4}$--4$_{1,3}$)}}
\newcommand{\hhco}{\mbox{H$_2$CO (3$_{0,3}$--2$_{0,2}$)}}
\newcommand{\hhhco}{\mbox{H$_2$CO (3$_{2,1}$--2$_{2,0}$)}}
\newcommand{\hhhhco}{\mbox{H$_2$CO (3$_{2,2}$--2$_{2,1}$)}}

\newcommand{\kms}{km\,s$^{-1}$}
\newcommand{\mjyb}{mJy\,beam$^{-1}$}

\newcommand{\umjyb}{$\mu$Jy\,beam$^{-1}$}

\begin{document}

\title{Early Planet Formation in Embedded
Disks (eDisk) XXII: \\ Keplerian disk, disk structures and jets/outflows in the Class 0 protostar IRAS 04166+2706}

\correspondingauthor{}
\email{Nguyen Thi Phuong: ntphuong02@vnsc.org.vn; Chang Won Lee: cwl@kasi.re.kr}

\author[0000-0002-4372-5509]{Nguyen Thi Phuong}
\affiliation{Vietnam National Space Center, Vietnam Academy of Science and Technology, 18 Hoang Quoc Viet, Nghia Do, Hanoi, Vietnam}
\affiliation{Korea Astronomy and Space Science Institute, 776 Daedeok-daero, Yuseong-gu, Daejeon 34055, Republic of Korea}
\email{ntphuong02@vnsc.org.vn}

\author[0000-0002-3179-6334]{Chang Won Lee}
\affiliation{Division of Astronomy and Space Science, University of Science and Technology, 217 Gajeong-ro, Yuseong-gu, Daejeon 34113, Republic of Korea}
\affiliation{Korea Astronomy and Space Science Institute, 776 Daedeok-daero, Yuseong-gu, Daejeon 34055, Republic of Korea}
\email{}

\author[0000-0002-6195-0152]{John J. Tobin}
\affil{National Radio Astronomy Observatory, 520 Edgemont Rd., Charlottesville, VA 22903 USA}
\email{}

\author[0000-0003-0998-5064]{Nagayoshi Ohashi}
\affiliation{Academia Sinica Institute of Astronomy and Astrophysics, 11F of Astronomy-Mathematics Building, AS/NTU, No.\ 1, Sec.\ 4, Roosevelt Rd, Taipei 10617, Taiwan, R.O.C}
\email{}

\author[0000-0001-9133-8047]{Jes K. J{\o}rgensen}
\affil{Niels Bohr Institute, University of Copenhagen, Jagtvej 155A, DK 2200 Copenhagen N., Denmark}
\email{}

\author[0000-0003-0845-128X]{Shigehisa Takakuwa}
\affiliation{Department of Physics and Astronomy, Graduate School of Science and Engineering, Kagoshima University, 1-21-35 Korimoto, Kagoshima, Kagoshima 890-0065, Japan}
\affiliation{Academia Sinica Institute of Astronomy and Astrophysics, 11F of Astronomy-Mathematics Building, AS/NTU, No.\ 1, Sec.\ 4, Roosevelt Rd, Taipei 10617, Taiwan, R.O.C}
\email{}

\author[0000-0003-3283-6884]{Yuri Aikawa}
\affiliation{Department of Astronomy, Graduate School of Science, The University of Tokyo, 7-3-1 Hongo, Bunkyo-ku, Tokyo 113-0033, Japan}
\email{}

\author[0000-0002-8238-7709]{Yusuke Aso}
\affiliation{Korea Astronomy and Space Science Institute, 776 Daedeok-daero, Yuseong-gu, Daejeon 34055, Republic of Korea}
\email{}

\author[0000-0002-7402-6487]{Zhi-Yun Li}
\affiliation{University of Virginia, 530 McCormick Rd., Charlottesville, Virginia 22904, USA}
\email{}

\author[0000-0003-2777-5861]{Patrick M. Koch}
\affil{Academia Sinica Institute of Astronomy and Astrophysics, 11F of Astronomy-Mathematics Building, AS/NTU, No.\ 1, Sec.\ 4, Roosevelt Rd, Taipei 10617, Taiwan, R.O.C}
\email{}

\author[0000-0001-5058-695X]{Jonathan P. Williams}
\affiliation{Institute for Astronomy, University of Hawai‘i at Mānoa, 2680 Woodlawn Dr., Honolulu, Hawai‘i 96822, USA}
\email{}

\author[0000-0001-5782-915X]{Sacha Gavino}
\affiliation{Niels Bohr Institute, University of Copenhagen, {\O}ster Voldgade 5--7, DK~1350 Copenhagen K., Denmark}
\email{}

\author[0000-0001-7233-4171]{Zhe-Yu Daniel Lin}
\affiliation{University of Virginia, 530 McCormick Rd., Charlottesville, Virginia 22904, USA}
\email{}

\author[0000-0001-8105-8113]{Kengo Tomida}
\affiliation{Astronomical Institute, Graduate School of Science, Tohoku University, Sendai 980-8578, Japan}
\email{}

\author[0000-0003-4022-4132]{Woojin Kwon}
\affiliation{Department of Earth Science Education, Seoul National University, 1 Gwanak-ro, Gwanak-gu, Seoul 08826, Republic of Korea}
\affiliation{SNU Astronomy Research Center, Seoul National University, 1 Gwanak-ro, Gwanak-gu, Seoul 08826, Republic of Korea}
\email{}

\author[0000-0002-4540-6587]{Leslie W.  Looney}\affiliation{Department of Astronomy, University of Illinois, 1002 West Green St, Urbana, IL 61801, USA}
\affil{National Radio Astronomy Observatory, 520 Edgemont Rd., Charlottesville, VA 22903 USA} 
\email{}

\author[0000-0002-9143-1433]{Ilseung Han}
\affiliation{Division of Astronomy and Space Science, University of Science and Technology, 217 Gajeong-ro, Yuseong-gu, Daejeon 34113, Republic of Korea}
\affiliation{Korea Astronomy and Space Science Institute, 776 Daedeok-daero, Yuseong-gu, Daejeon 34055, Republic of Korea}
\email{}

\author[0000-0001-6267-2820]{Alejandro Santamaría-Miranda}
\affiliation{European Southern Observatory, Alonso de Cordova 3107, Casilla 19, Vitacura, Santiago, Chile}
\email{}

\author[0000-0001-5522-486X]{Shih-Ping Lai}
\affiliation{Institute of Astronomy, National Tsing Hua University, No. 101, Section 2, Kuang-Fu Road, Hsinchu 30013, Taiwan}
\affiliation{Center for Informatics and Computation in Astronomy, National Tsing Hua University, No. 101, Section 2, Kuang-Fu Road, Hsinchu 30013, Taiwan}
\affiliation{Department of Physics, National Tsing Hua University, No. 101, Section 2, Kuang-Fu Road, Hsinchu 30013, Taiwan}
\affiliation{Academia Sinica Institute of Astronomy and Astrophysics, 11F of Astronomy-Mathematics Building, AS/NTU, No.\ 1, Sec.\ 4, Roosevelt Rd, Taipei 10617, Taiwan, R.O.C}
\email{}

\author[0000-0003-1412-893X]{Yen Hsi-Wei}
\affil{Academia Sinica Institute of Astronomy and Astrophysics, 11F of Astronomy-Mathematics Building, AS/NTU, No.\ 1, Sec.\ 4, Roosevelt Rd, Taipei 10617, Taiwan, R.O.C}
\email{}

\author[0000-0003-0334-1583]{Travis J. Thieme}
\affiliation{Institute of Astronomy, National Tsing Hua University, No. 101, Section 2, Kuang-Fu Road, Hsinchu 30013, Taiwan}
\affiliation{Center for Informatics and Computation in Astronomy, National Tsing Hua University, No. 101, Section 2, Kuang-Fu Road, Hsinchu 30013, Taiwan}
\affiliation{Department of Physics, National Tsing Hua University, No. 101, Section 2, Kuang-Fu Road, Hsinchu 30013, Taiwan}
\email{}

\author[0000-0003-4361-5577]{Jinshi Sai (Insa Choi)}
\affiliation{Academia Sinica Institute of Astronomy and Astrophysics, 11F of Astronomy-Mathematics Building, AS/NTU, No.\ 1, Sec.\ 4, Roosevelt Rd, Taipei 10617, Taiwan, R.O.C}
\email{}

\author[0000-0002-8591-472X]{Christian Flores}
\affiliation{Academia Sinica Institute of Astronomy \& Astrophysics, 11F of Astronomy-Mathematics Building, AS/NTU, No.1, Sec. 4, Roosevelt Rd,Taipei 10617, Taiwan, R.O.C.}
\email{}

\begin{abstract}
We present ALMA observations of the Class 0 protostar IRAS 04166+2706, obtained as part of the ALMA large program Early Planet Formation in Embedded Disks (eDisk). These observations were made in the 1.3\,mm dust continuum and molecular lines at angular resolutions of $\sim 0.05''$ ($\sim 8$\,au) and $\sim 0.16''$ ($\sim25$\,au), respectively. The continuum emission shows a disk-like structure with a radius of $\sim22$\,au.  Kinematical analysis of \mbox{\tco}, \cho, \hhco, \mbox{\metanol}~emission demonstrates that these molecular lines trace the infalling-rotating envelope and possibly a Keplerian disk, enabling us to estimate the protostar mass to be $0.15\,\rm{M_\odot} < \rm{M_\star} < 0.39 \,M_\odot$.
The dusty disk is found to exhibit a brightness asymmetry along its minor axis in the continuum emission, probably caused by a flared distribution of the dust and the high optical depth of the dust emission. In addition, the \co~and \sio~emissions show knotty and wiggling motions in the jets. Our high angular resolution observations revealed the most recent mass ejection events, which have occurred within the last $\sim 25$ years. 

\end{abstract}
\keywords{stars: protostars, stars: jets, stars: individual (IRAS 04166 +2706)}

\section {Introduction} 
\label{sec:intro}
Planets are expected to form in circumstellar disks surrounding protostars. Therefore, studying the gas and dust properties in disks is essential to understand the formation processes of planetary systems. Observations of protoplanetary disks revealed prominent substructures, including ring, gap, crescent, and spiral. The formation of these features is attributed to several proposed mechanisms such as planet-disk interactions \citep{Paardekooper+Mellema_2004}, photoevaporation \citep{Alexander+Clarke+etal-2006}, magneto-rotational and hydrodynamic instabilities \citep{Flock+Ruge+etal_2015}, dust accumulation at snowlines \citep{Zhang+Jin+2015}, gravitational or stellar perturbations \citep{Dong+Hall_2015}, and the processes of dust evolution and radial drift \citep{Ohashi-S+etal_2021}. Among these proposed hypotheses, planet-disk interaction remains the leading interpretation for the observed ring and gap patterns. Indeed, ring and gap structures that can be precursors of planet formation have been observed in Class I protostars such as WL 17 and GY 91 \citep{Sheehan+etal_2017, Sheehan+etal_2018}, \mbox{IRS 63} \citep{Cox+etal_2020}, in a flat-spectrum protostar, BHB[2007]1 \citep{Alves+etal_2020}, and in 7 Class 0/I/Flat spectrum disks in Orion \citep{Sheehan+etal_2020}. 
This supports the hypothesis that planetesimals and ultimately giant planet cores are formed in the disk during the earlier stages of the evolution of the protostar when it is still deeply embedded within its natal envelope \citep{Greaves+Rice_2010, Booth+Clarke_2016}.  

To investigate the possibility of planet formation during the protostellar phase, 12 Class 0 and 7 Class I protostars were observed as part of the ALMA Large Program Early Planet Formation in Embedded Disks \citep[eDisk,][]{Ohashi+edisk_2023} in dust continuum and line emission at the angular resolutions of 0.03 -- 0.1\arcsec. The main objective of the eDisk program is to search for possible primordial signature of planet formation, the substructures in the disks surrounding these young protostars.

In this paper, we present observations of \mbox{IRAS 04166+2706}, a young stellar object \citep{Chen+etal_1995, Park+etal_2002} located in the B213 cloud, part of the Taurus molecular cloud complex. The distance to B213 was estimated to be 161.3 pc by \citet{Luhman_2018} based on Gaia DR2 measurements of stars associated with the cloud. 
\citet{Galli+Loinard+etal_2019} used both Gaia DR2 and VLBI data to estimate the distance of the cloud B213 to be 160.7\,pc (+2.6/--2.5).
\citet{Roccatagliata+etal_2020} classified B213 as part of Taurus E and reported the distance of Taurus E based on Gaia DR2 data of 160.2 (+/-0.6)\,pc. 
More recently, \citet{Krolikowski+etal_2021} used the new data of Gaia EDR3 and calculated the distance of the region to be 155.94\,pc. 
For this paper, we adopt the most recent distance of the source as 156\,pc. Using the revised distance as well as the spectral energy distribution from infrared through millimeter wavelengths, the eDisk collaboration estimated the bolometric luminosity and temperature of the source to be 0.41\,L$_\odot$ and 54\,K respectively, classifying it as a Class 0 protostar \citep{Ohashi+edisk_2023}. The youth of the source is also witnessed by its very high-velocity outflows and collimated jets as reported by single dish and interferometric observations \citep{Tafalla+Santiago+etal_2004, Tafalla+Su+etal_2017, Santiago+etal_2009, Wang+Shang+etal_2014, Wang+Shang+etal_2019}. 

This paper is laid out as follows: the observations and data reduction are described in Section \ref{sec:obs}. Section \ref{sec:obs_results} introduces the observing results of the 1.3\,mm dust and the molecular line (CO, $^{13}$CO, C$^{18}$O, H$_2$CO, and CH$_3$OH) emission toward IRAS 04166+2706. Section \ref{sec:ana} presents the data analysis, where we explore the dust disk morphology, and its possible substructure, as well as the Position-Velocity diagrams of \tco, \cho, \hhco~ and \metanol~ line emission in search for Keplerian gas motion in the disk to estimate the dynamical mass of the central protostar. We discuss the results in Section \ref{sec:dis} and provide a summary of the paper in Section \ref{sec:sum}.

\section{Observation and Data Reduction}\label{sec:obs}
\begin{table*}
\centering 
\renewcommand{\arraystretch}{1.35}
\setlength{\tabcolsep}{1.em}
\caption{Summary of the ALMA observations of IRAS 04166+2706}\label{tab:obs}
\begin{tabular}{ccccc}
\hline\hline 
Date & \multicolumn{2}{c}{30 September, 1, 18, 24} October 2021 & \multicolumn{2}{c}{ 3 July, 2022}\\
Configuration & \multicolumn{2}{c}{C43--8} &  \multicolumn{2}{c}{C43--5}   \\
Projected baseline (m) & \multicolumn{2}{c}{46--11886} & \multicolumn{2}{c}{15--1301}\\
Time on source & \multicolumn{2}{c}{3.0 hours} & \multicolumn{2}{c}{57  minutes} \\
Calibrator & \multicolumn{2}{c}{J0328+1636, J0433+2905}  & \multicolumn{2}{c}{J0510+1800, J0438+3004}  \\
Phase center  & \multicolumn{2}{c}{$04^{\rm{h}}19^{\rm{m}}42\fs51$, $+27^{\circ}13{\arcmin}35\farcs8$}& \multicolumn{2}{c}{$04^{\rm{h}}19^{\rm{m}}42\fs51$, $+27^{\circ}13{\arcmin}35\farcs8$} \\
\hline
    & Continuum$^\dag$ & \co$^{\dag\dag}$ & $\tco^{\dag\dag}$ & $\cho^{\dag\dag}$ \\
Frequency (GHz) & 225 & 230.53800  & 220.39868 & 219.56035  \\
Freq. width/vel. resolution & $\sim$2\,GHz & 0.635\,\kms & 0.167\,\kms & 0.167\,\kms \\ 
E$_u$ (K) & & 16.6 & 15.9 & 15.8 \\
Beam (P.A.)  &$0.05''\times0.04''$(22$^\circ$)& $0.17''\times0.14''$(7$^\circ$)  & $0.18''\times0.15''$(6$^\circ$) & $0.18''\times0.14''$(8$^\circ$) \\ 
rms  & 23\,\umjyb & 2.0\,\mjyb & 3.1\,\mjyb& 2.4\,\mjyb\\
\hline
& & \sio$^{\dag\dag}$ & \metanol$^{\dag\dag}$ & \hhco$^{\dag\dag}$ \\ 
Frequency (GHz) & & 217.10498 & 218.4401 & 218.2222 \\
E$_u$ (K) & & 31.3 & 45.5 & 21.0 \\
Vel. resolution &  & 1.34\,\kms & 1.34\,\kms & 1.34\,\kms \\ 
Beam (P.A.) & &$0.18''\times0.14''$(6$^\circ$) & $0.17''\times0.14''$(6$^\circ$) & $0.17''\times0.14''$(6$^\circ$)\\ 
rms & & 1.0\,\mjyb & 0.7\,\mjyb & 0.8\,\mjyb\\
\hline\hline
\end{tabular} 

\footnotesize{$^{\dag}:$ robust=0.0, $^{\dag\dag}:$ robuts=2.0}
\end{table*}

IRAS 04166+2706 was observed as part of the ALMA Large program eDisk: Early Planet Formation in Embedded Disks \citep[2019.1.00261.L, \textit{PI: Nagayoshi Ohashi}, see also][for an overview of the program]{Ohashi+edisk_2023} using two configurations C43-8 and C43-5. 
The long baseline data (configuration C43-8) were obtained in 4 execution blocks (EBs) on 2021 September 30, October 1, 18, and 24, providing an observing time of $\sim 6.7$ hours with a total on-source time of $\sim 3.0$ hours. The numbers of antennas used in each block were 44, 45, 43, and 45, respectively. The baselines among the antennas ranged from 15\,m to 11.89\,km.

The short baseline data (configuration C43-5) were obtained in 2 blocks on 2022 July 3 with a total observing time of 2.95\,hours and on-source time of \mbox{$\sim 57$\,minutes}. The number of effective antennas used in these blocks is 41, resulting in the baselines among the antennas from 15\,m to 1.3\,km. 
The details of the spectral setup and the procedures for data processing can be found in the overview paper of eDisk \citep{Ohashi+edisk_2023}.  Here, we briefly summarize their main points only. 

The data were reduced using the CASA version 6.2.1 and pipeline version 2021.2.0.128. The three latter EBs of long baseline data have the QA0 status of ``semi-pass''\footnote{QA stands for Quality-Assurance. The QA0 is a near-real-time verification of data quality, reflecting the quality of the calibrations included in each EB. The QA2 includes full calibration of the entire collection of EBs.} due to high phase decoherence. We downloaded the raw data and ran the ALMA calibration pipeline manually. The continuum emission was obtained from line-free channels. We then imaged each execution block (EB) individually and shifted the brightest source in the image to common phase center which was obtained from the \textit{imfit} task in CASA. After that, we computed the azimuthally averaged visibility amplitude of each dataset and plotted them together to determine the average scale factors between each EB. At this stage, we realized that the amplitude profiles of these EBs exhibited different behaviors. Therefore, we could not apply the standard eDisk reduction procedure to this dataset but the ``two-pass'' method is needed \citep[see also][]{Ohashi+edisk_2023}. Namely, we first completed the self-calibration process and determined the scale factors after self-calibration. We then reran the entire procedure from the beginning, applying these scale factors a priori. This process was performed first with the short baseline data only and then with the short baseline and long baseline data combined. Both phase self-calibration and amplitude self-calibration were used\footnote{See detail in https://github.com/jjtobin/edisk}. The self-calibration solutions were then applied to all line emission.

The visibility data were imaged with different robust parameters using the task \textit{tclean} in CASA. For this paper, we present the image of dust emission produced with a robust parameter of 0.0, while all the lines have been produced with a robust parameter of 2.0. All the maps are corrected for primary beam attenuation, and the noise levels of the line maps were measured in emission-free channels. To emphasize the presence of extended dust emission, we also present the dust emission image produced using a robust parameter of 2.0 and an $uv$ taper of 1000$k \lambda$. The main analyses of the continuum emission in this paper are performed using the image with a robust parameter 0.0 which is found to provide the best compromise between the angular resolution and sensitivity. Table 1 summarizes the main parameters of our ALMA observations for IRAS 04166+2706.

\section{Results}\label{sec:obs_results}
\subsection{Dust emission}
\begin{figure*}
\centering
\includegraphics[width=0.8\linewidth]{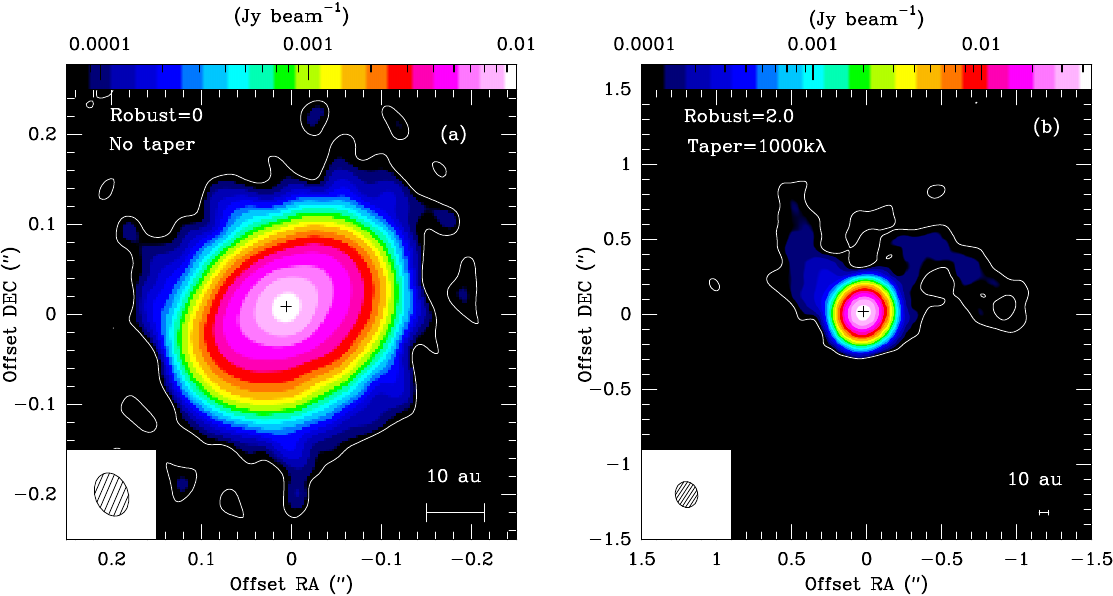}
\caption{\textit{Left:} Image of the 1.3\,mm continuum emission produced with a robust parameter of 0. The color scale is in the unit of Jy\,beam$^{-1}$ and the beam with a size of $0.05''\times0.04''\, \rm{and} \, \rm{P.A.=22^\circ}$ is shown in the lower left corner. \textit{Right:} The same image but with the robust parameter of 2.0 and a taper of 1000$k\lambda$. The beam has a size of $0.20''\times0.17''\, \rm{and} \, \rm{P.A.}=8.0^\circ$ in this case. The contour is at $3\sigma$ level, with $1\sigma$ values of 23\, \umjyb~in the left panel and 33\,\umjyb~in the right panel. The black cross shows the central position as derived from the 2D Gaussian fit.}\label{fig:cont}
\end{figure*} 

Figure \ref{fig:cont} shows the intensity maps of 1.3\,mm dust emission produced using \textit{robust=0.0} (left) and \textit{robust=2.0} with a taper of 1000 $k \lambda$ (right). The left panel shows the disk-like structure out to 0.2$''$ in radius with the peak emission slightly shifted to the northeast direction from the map center (phase center). The right panel shows that besides the compact disk-like structure in the center, there are also two faint arm-like features extending out to $\sim 1.0''$.

We fit the continuum image produced with the \mbox{\textit{robust=0.0}} with a model of 2D Gaussian using \textit{imfit} task in CASA. The results of the 2D Gaussian fit are given in Table \ref{tab:imfit}. The deconvolved size of the dusty disk is \mbox{$138\,\rm{mas} \times 94\,\rm{mas}$}, which corresponds to a linear size of  $\sim$22\,au$\times$15\,au (at the distance of 156\,pc). The position angle of the disk is P.A.=$122^\circ$ (measured counterclockwise from the north), and the inclination of the disk is $\sim47^\circ$ from $\cos^{-1}(94/138) $ under the assumption of a vertically thin circular disk geometry. The flux density of the continuum emission is $70.8\pm0.3$\,mJy. We note that the fit was performed with the image produced using \textit{robust=0.0} without any $uv$ tapering. So, possibly some extended emission (e.g. the emission from the weak envelope) may not be included in this fit value. 
Indeed, when all the emission greater than $3\sigma$ in the image produced with a robust value of 2.0 and a $uv$ taper of 1000$k \lambda$ (Figure \ref{fig:cont}b) are summed up, the total flux density increases to 73.5$\pm$0.15\,mJy.
 Note that the total flux density that we detected is $\sim20$\% higher than the value of $59\pm2$\,mJy by IRAM PdBI observations at 1.3\,mm \citep{Santiago+etal_2009}.

\begin{table}[ht!]
\centering
\caption{Result of the 2D Gaussian fit to dust continuum image of \textit{robust=0.0}}\label{tab:imfit}
\centering 
\renewcommand{\arraystretch}{1.15}
\setlength{\tabcolsep}{1.2em}
\begin{tabular}{cc}
\hline\hline
\multirow{2}{*}{Position (ICRS)} & 
R.A: $04^{\rm{h}}19^{\rm{m}}42\fs51$ \\
& DEC: $+27{^\circ}13\arcmin35\farcs8$ \\
Major axis & $137.9\pm0.7$ mas\\ 
Minor axis & $93.7\pm0.5$ mas\\ 
Position angle & $121.5^\circ\pm0.5^\circ$\\
Flux density & $70.8\pm0.3$\,mJy\\
\hline\hline
\end{tabular}

\footnotesize{The major axis, minor axis and position angle are deconvolved}
\end{table} 

\subsection{Molecular line emission}
\begin{figure*}
 \centering
     \includegraphics[width=0.95\linewidth]{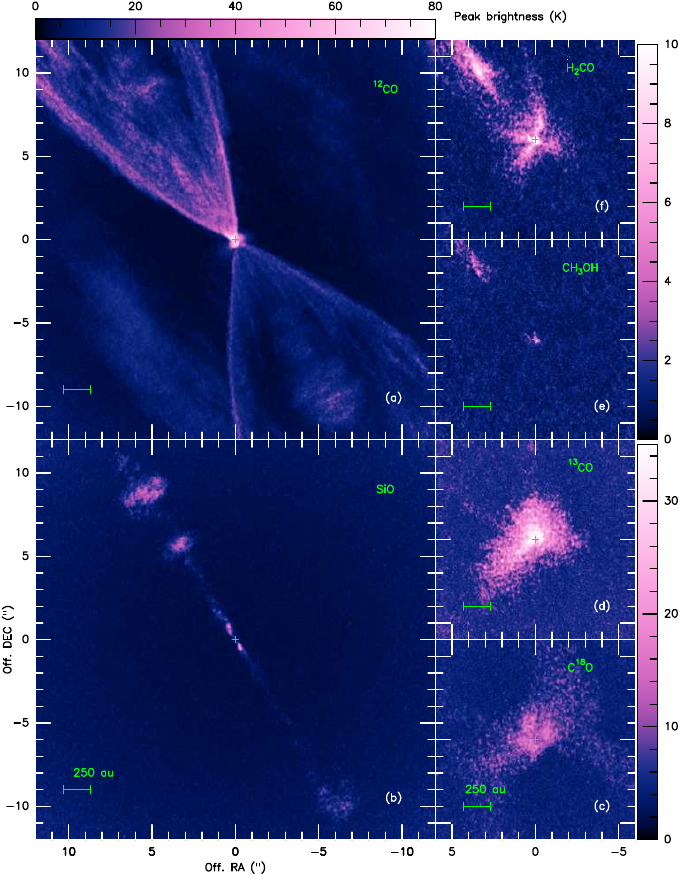}
    \caption{Peak brightness temperature maps of \co (a),  \sio (b), \cho (c), \tco (d), \metanol (e),  and \hhco (f)  emission.
    The green cross marks the central position of the 1.3\,mm continuum emission, derived from the 2D Gaussian fit (Table \ref{tab:imfit}).
     The green bar in the lower left corner of each panel indicates a linear scale of 250\,au ($\sim$10 times the beam size) at a distance of 156\,pc. The \sio~and \co~maps,  the \hhco~and \metanol~maps, and the \tco~and \cho~maps are drawn according to the color bar scales on the top of the \co~map, on the right of two upper right panels, and on the right of two lower right panels, respectively.}
    \label{fig:Tpeak-all}
\end{figure*}

 Figure \ref{fig:Tpeak-all} shows the peak brightness temperature maps of \co, \tco, \cho, \mbox{\hhco}, and \metanol. The brightness temperature was computed using Planck function. We describe detailed features of each molecular line emission below.
 
\begin{figure*}[htbp!]
\centering
    \includegraphics[width=0.265\linewidth]{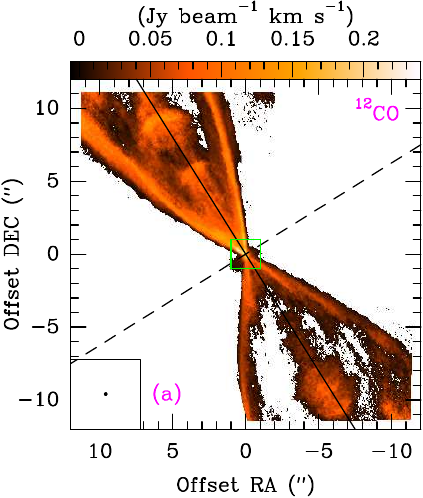}
    \includegraphics[width=0.45\linewidth]{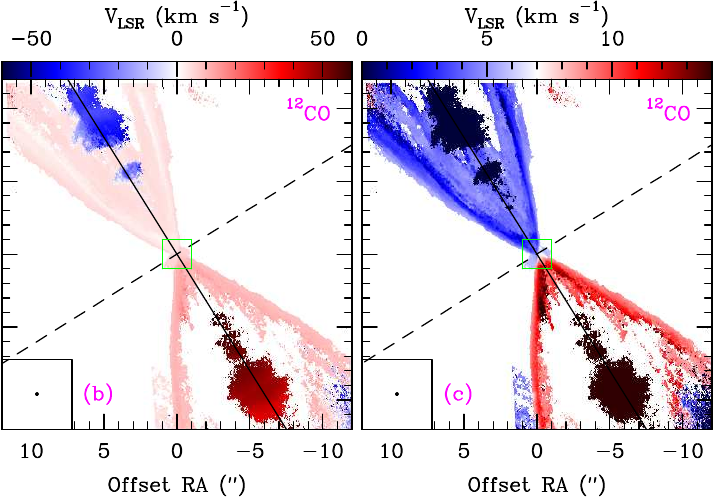}
    \includegraphics[width=0.265\linewidth]{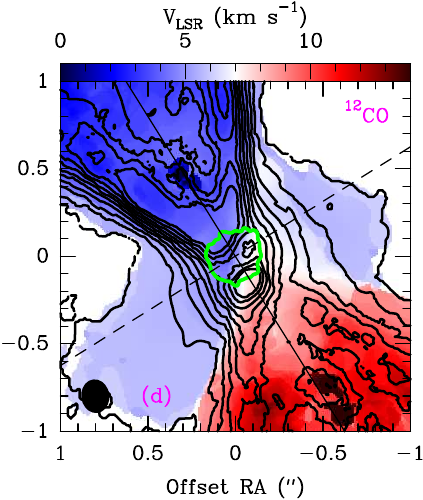}
    \caption{Moment 0 (a) and moment 1 (b, c) maps of the \co~emission. The moment 1 maps were generated at different color scales to highlight the high-velocity jets (b) and the bipolar conical outflows at a lower velocity (c). The color scale is indicated on the top of each panel in units of Jy\,beam$^{-1}$\,km\,s$^{-1}$ (a) and \kms (b, c),   respectively. (d): Zoomed-in moment 0 (contours) and moment 1 (colors) maps of the central region that marked in green square in the left panels. Contour levels start at 3$\sigma$ and then increase in steps of 3$\sigma$. The green contour in the lower right panel indicates continuum emission (robust=0.0) at $5\sigma$. In all panels, the beam of \mbox{$0.17''\times0.14''$ (PA=7$^\circ$)} is indicated in the lower left corner, and directions of the major (PA=121.5$^\circ$) and minor (PA=31.5$^\circ$) axes of the dust emission are shown by the dashed and solid black lines, respectively.}
    \label{fig:12co-moments}
\end{figure*}

\textbf{\co~and \sio:} Figure \ref{fig:12co-moments} shows the integrated intensity (moment 0) and mean velocity \mbox{(moment 1)} maps of the \co~emission. Both moment 0 and 1 maps were made with a threshold of $3\sigma$ in the velocity range of \mbox{$V_{LSR}=-60~\rm{to}~ 60$\,\kms}. 
The panels (b) and (c) show mean velocity maps of the high- and low- velocity components, respectively. These maps display the conical outflows and jets structures where the blue-shifted emission is stronger on the north-east side than the red-shifted emission, as shown in the peak brightness temperature map (Figure \ref{fig:Tpeak-all}), which was also reported by \citet{Santiago+etal_2009}, \citet{Wang+Shang+etal_2014} and \citet{Tafalla+Su+etal_2017}. The structures of conical outflows and jets are almost symmetric with respect to the minor axis of the dust disk, suggesting that the the axis of the outflows and jets are almost aligned with the disk minor axis. 
We note that there is also \co~emission distribution perpendicular to the outflows axis around the protostar, as better seen in Figure \ref{fig:12co-moments}(d), which may mostly originate from the disk/envelope. 

\begin{figure*}[htbp!]
    \centering  
    \includegraphics[width=0.334\linewidth]{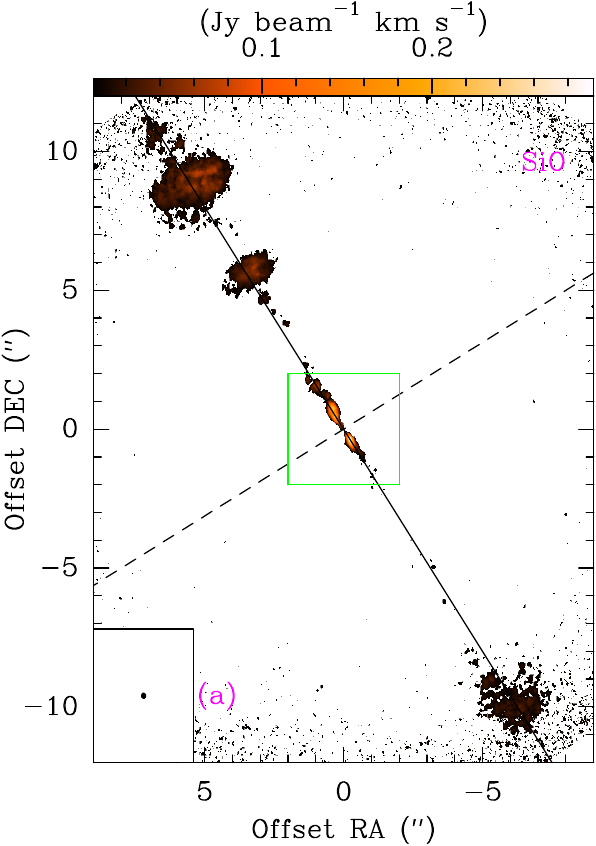}
    \includegraphics[width=0.283\linewidth]{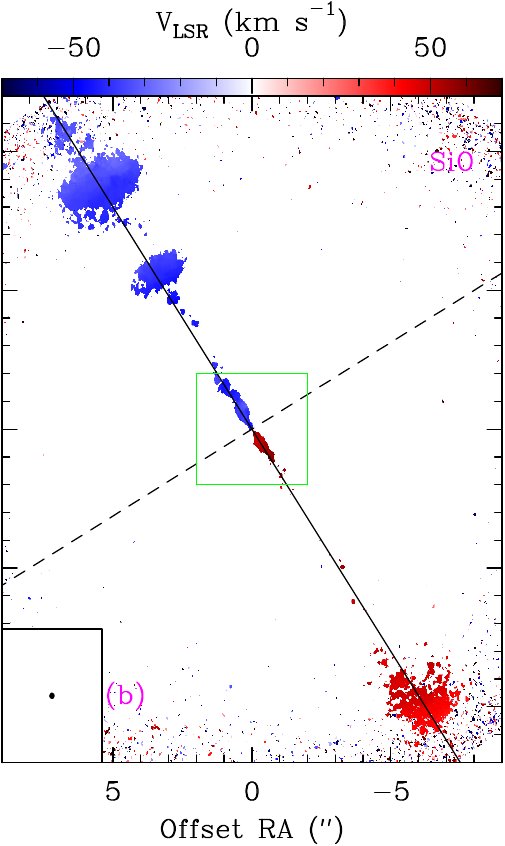}
    \includegraphics[width=0.338\linewidth]{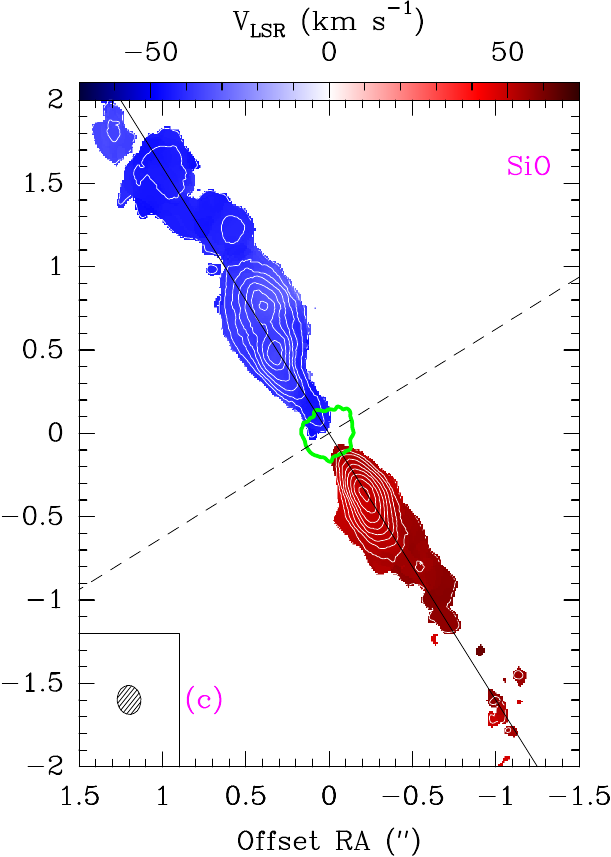}
    \caption{Moment 0 (a) and moment 1 (b) maps of \sio~emission. The color scale on the top of each panel is in units of Jy\,beam$^{-1}$\,km\,s$^{-1}$ and \kms, respectively. (c): Zoomed-in moment 0 (contours) and moment 1 (colors) maps of the central square region marked in green in the left panels. Contour levels start at 3$\sigma$ and then increase in steps of 6$\sigma$. The green contour is continuum emission (robust=0.0) at $5\sigma$. In all panels, the beam of \mbox{$0.18''\times0.14''$ (PA=6$^\circ$)} is indicated in the lower left corner, and directions of the major (PA=121.5$^\circ$) and minor (PA=31.5$^\circ$) axes of the dust emission are shown by the dashed and solid black lines, respectively.}
    \label{fig:sio-moments}
\end{figure*}

Figure \ref{fig:sio-moments} shows the moment maps of \sio. The moments were computed in a velocity range of --70 to 70\,\kms~ with a threshold of 3\,$\sigma$. The \sio~emission reveals episodic jets at a high velocity of \mbox{$|V_{LSR}|>20$ \kms}, with the blue-shifted emission stronger than the red-shifted counterpart. The right panel of this figure shows a closer view of the central region, which reveals several intensity peaks (contours)
and a stretched S-shape morphology on the blue-shifted side, suggesting wiggling or precession motions in the jet. Intensity maps integrated in narrower velocity ranges will be discussed in Section \ref{sec:jets_outflows}.

\begin{figure*}
\centering
\includegraphics[width=0.33\linewidth]{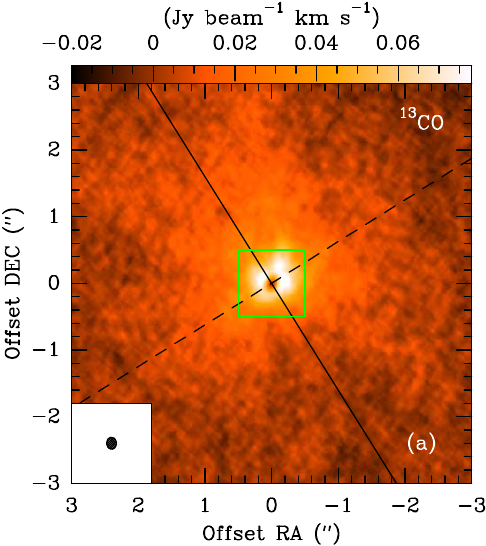}
\includegraphics[width=0.6\linewidth]{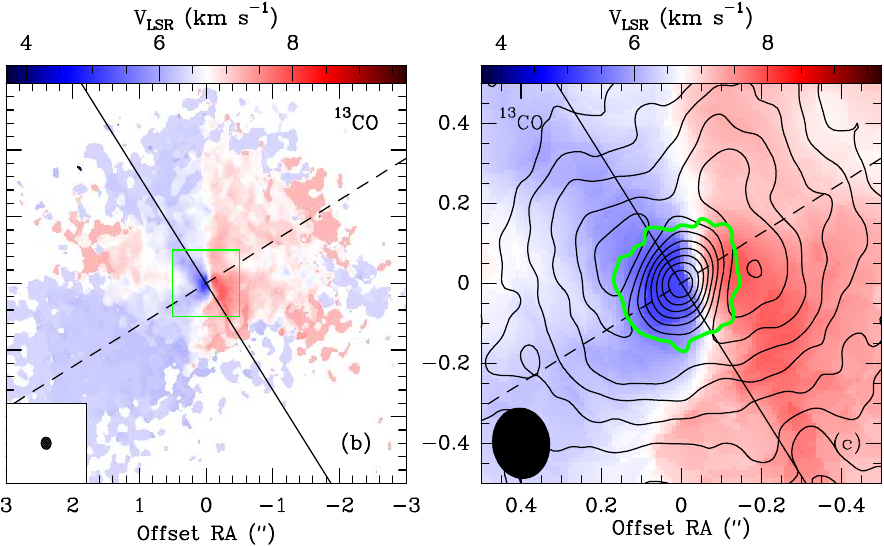}
\vspace{0.5cm}
\includegraphics[width=0.33\linewidth]{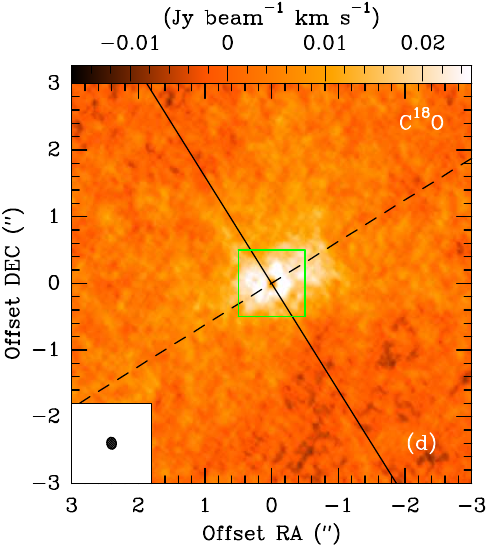}
\includegraphics[width=0.6\linewidth]{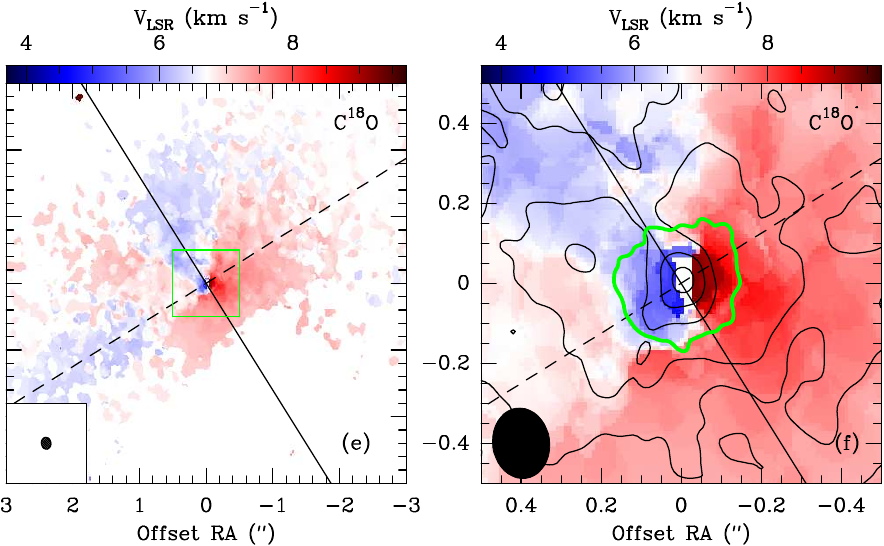}
\caption{\textit{Upper panels:} \tco~emission. Maps of moment 0 (a), moment 1 (b), and moment 0 (contours) and moment 1 (colors) maps zoom-in in the green box indicated in the left panels (c). The color scales on the top of each panel are in units of Jy\,beam$^{-1}$\,km\,s$^{-1}$ and \kms~, respectively. In panel (c), the contour levels start at 3$\sigma$ and then increase in steps of 3$\sigma$. The green contour is $5\sigma$ level of dust emission (robust=0.0). In all panels, the beam of \mbox{$0.18''\times0.15''$ (PA=6$^\circ$)} is indicated in the lower left corner. The dashed and solid black lines denote the major (PA=121.5$^\circ$) and minor (PA=31.5$^\circ$) axes of the disk structure in the dust emission, respectively. \textit{Lower panels:} Same as the upper panels but for \cho. The beam size is $0.18''\times0.14''$ (PA=$8^\circ$).} \label{fig:c18o+13co_moments}
\end{figure*}

\textbf{\tco~and \cho:} 
Figure \ref{fig:c18o+13co_moments} shows moment 0 and 1 maps of \tco~and \cho~emission. The moment 0 maps are made in the velocity range of $V_{LSR}$=3.5 to 10\,\kms, and the velocity maps are produced with a threshold of 4\,$\sigma$. Note that both intensities of \tco~and \cho~emission are significantly dropped at the center. This may be caused by the absorption of line emission by the continuum. The velocity maps show complicated gas kinematics with red-shifted and blue-shifted emission mixed, probably revealing the combination of the disk and envelope kinematics. The right panels of this figure show the same velocity maps zoomed in the central protostar region. The dust disk region shows a velocity gradient between red-shifted emission on the western side and blue-shifted emission on the eastern side. The direction of the velocity gradient appears to be misaligned from the direction of the major axis of the dust disk. The blue-shifted emission close to the central star may trace some part of the outflow, as such blue-shifted emission is also well distributed toward the north-east side. However, as shown in the Figure \ref{fig:c18o+13co_moments} (c) and (f), the features are very close to the area where dust emission detected, suggesting that velocity pattern can be also interpreted as a mixture of rotation and infalling motion in the flattened disk/envelope \citep[e.g.,][]{Momose_1998}.

\begin{figure*}
\centering
\includegraphics[width=0.33\linewidth]{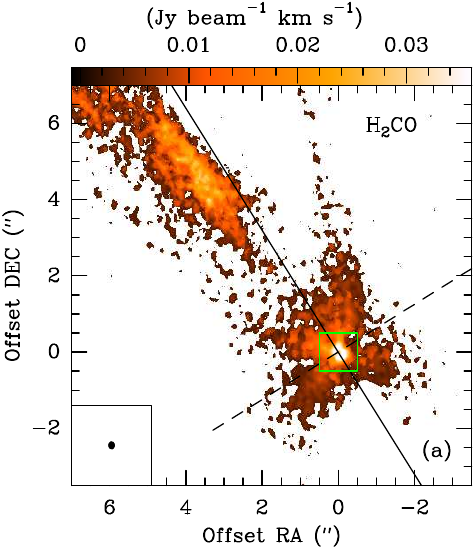}
\includegraphics[width=0.615\linewidth]{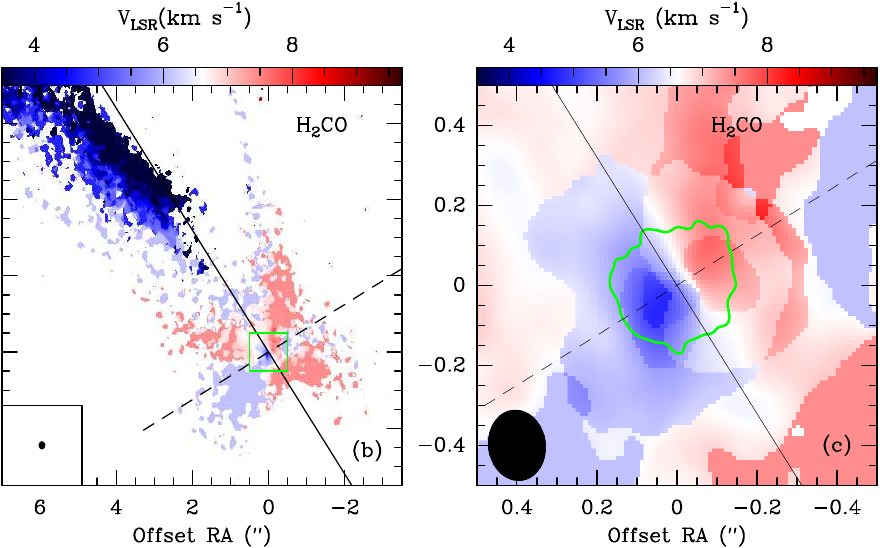}\\
\vspace{0.5cm}
\includegraphics[width=0.33\linewidth]{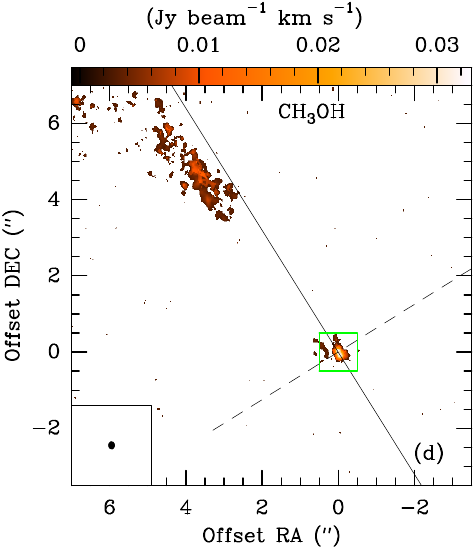}
\includegraphics[width=0.615\linewidth]{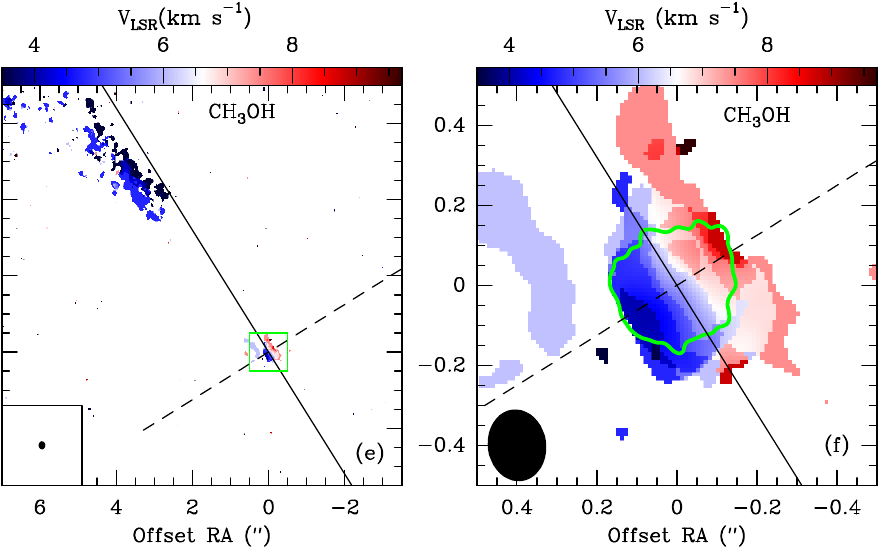}
\caption{\textit{Upper panels:} Moment maps of the \hhco~emission. Moment 0 (a), moment 1 (b), and zoomed-in moment 1 map (c) of the central region (marked as green boxes in the two left panels). In the panel (c), the green contour is $5\sigma$ level of dust emission (robust=0.0). In all panels, the beam of $0.17''\times0.14''$ (PA=6$^\circ$) is indicated in the lower-left corner of each panel, and the dashed and solid black lines show the direction of the major (PA=121.5$^\circ$) and minor (PA=31.5$^\circ$) axes of the dust emission, respectively. \textit{Lower panels:} Same as the upper panels but for \metanol~emission.} \label{fig:h2co+ch3oh-moments}
\end{figure*}

\textbf{\hhco~and \metanol:} 
Figure \ref{fig:h2co+ch3oh-moments} shows moment 0 and 1 maps of the \mbox{\hhco} and \metanol~emission. The moments were computed in the velocity range of $V_{LSR}$=3.5 to 10\,\kms~with a threshold of $3\sigma$.

The \hhco~maps show primarily two emission components; one component surrounding the dust disk, similar to the \tco~and \cho~emissions, and the other tracing the outflow-related components. The former one exhibits rather complicated velocity structures, which may be a combined effect of rotation, infall, and innermost outflowing gas motions. 
Zooming in the dust disk region, we see a clear velocity gradient along the major axis of the dust disk indicating the presence of rotating motions in the disk.

It is noted that this axis is slightly offset towards the blue-shifted emission. Although it is not clear why there is such an offset, our low spectral resolution \mbox{($\sim1.34$\,\kms)} observations of these lines can be one possible reason. 
The \hhco~maps include the blue-shifted outflow component to the north along the outflow axis, which is slightly tilted toward the southeast with respect to the outflow axis, and the X-shape outflow components near the protostar. Note that there is no red-shifted \hhco~emission along the outflow axis to the south, while there are blue and red-shifted components in the X-shape outflow.

The \metanol~emission is seen in the central region of the system, mostly coinciding with the dust emission, and has a velocity gradient along the major axis of the dust disk, again possibly tracing the rotational motions. In addition, the northeast blue-shifted outflow-related emission is also observed in \metanol~at about the same position and the same velocity as the \hhco~emission. Like the case of \hhco~emission, there is no counterpart of the outflow to the southwest direction. 

We present channel maps of all the mentioned lines in Appendix \ref{sec:channel-maps}. Our observations also include two other lines of H$_2$CO (\hhhco~and \hhhhco), \so, \dcn, and 3 lines of c-C$_3$H$_2$ (\Cyclopropenyo, \Cyclopropenyt, \Cyclopropenytt). We present their moment 0 and moment 1 maps in Appendix \ref{sec:other_lines}.
 
\clearpage
\section{Analysis} \label{sec:ana}
\subsection{Disk mass} 
\label{sub:disk_mass}
Assuming that the dust emission is optically thin and isothermal, we can estimate the disk dust mass using the following equation: 
\begin{equation} 
\centering
M_d=\frac{S_\nu\,d^2}{\kappa_\nu\,B_\nu(T_d)}, 
\end{equation}
where $\nu$ is the observed frequency (225\,GHz), $S_\nu$ is the flux density of dust emission (70.8\,mJy) derived from the previous 2D Gaussian fit (to the \textit{robust=0.0} image), $d$ is the source distance (156\,pc), $B_\nu(T_d)$ is the Planck function at the dust temperature of $T_d$, and $\kappa_\nu$ is the dust mass opacity at the observed frequency. We adopt the dust opacity of $\kappa_{1.3\,mm}$=2.3\,cm$^2$\,g$^{-1}$ \citep{Beckwith+etal_1990}, and assume a dust temperature of 20\,K which is the median value of dust temperatures of disks in the Taurus star-forming region as derived by \citet{Andrews+etal_2005}. Assuming a gas-to-dust ratio of 100, we obtained the disk mass of 0.015\,M$_{\odot}$. 

\citet{Tobin+etal_2020} introduced a more realistic method to estimate the dust temperature of Class 0/I objects based on the source luminosity as \mbox{$T_{\rm{d}}=43\,(L_{\rm{bol}}/\rm{L}_\odot)^{0.25}$}. Applying this equation to the case of IRAS 04166+2706 ($L_{\rm{bol}}=0.41\,\rm{L}_\odot$) we obtain the dust temperature of $\sim$34\,K in this disk. Using the dust temperature of 34\,K and all the other parameters the same as above, we obtain the total mass in the disk of 0.008\,M$_\odot$.  

\citet{Santiago+etal_2009} estimated a value of 0.02\,M$_\odot$ for the disk mass. They derived the mass from IRAM PdBI observations of 1.3\,mm using dust temperature of 20\,K and dust opacity of 1.0 \,cm$^{2}$\,g$^{-1}$ \citep{Ossenkopf+Henning_1994}. Adopting their value of opacity, we obtained 0.028\,M$_\odot$ for the disk mass, in good agreement with the value quoted by \citet{Santiago+etal_2009}. These estimates would be only valid if the emission is indeed optically thin. If it is optically thick, the derived mass would be a lower limit. 
We also emphasize that our estimate of the disk mass is also consistent with the estimate of (0.027\,M$_\odot$) from RADMC-3D modeling and MCMC fit using a wide range SED set data including CARMA mm data \citep{Sheehan+Eisner_2017}.   

\subsection{Identifying a possible substructure in the continuum disk}
\label{sec:dust-morp}
In this section, we attempt to identify any possible faint substructures in the continuum disk which can be a signature of primordial formation of proto-planets. 

\begin{figure}[htbp!]
    \centering
     \includegraphics[width=0.9\linewidth]{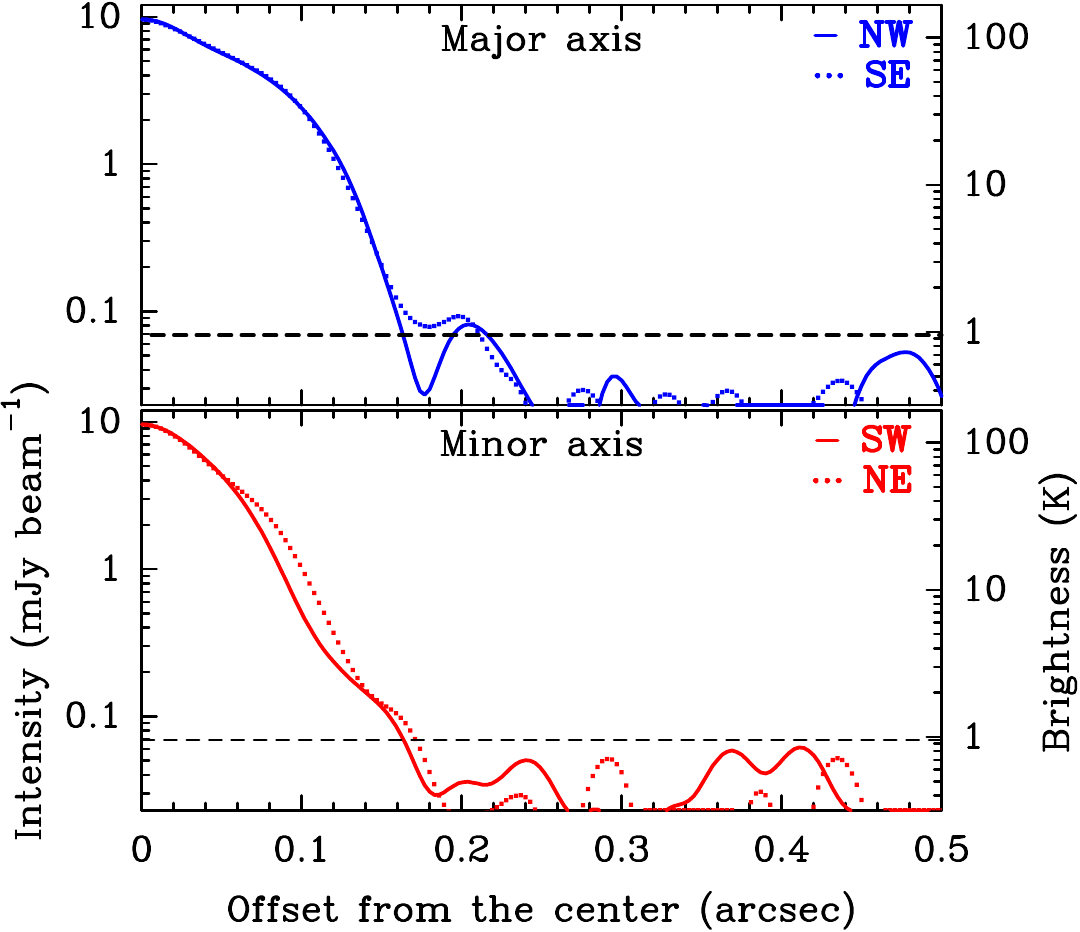}
    \caption{1.3\,mm dust emission profiles along the major (\textit{upper}) and minor (\textit{lower}) axes of the disk. The horizontal dash line indicates the $3\sigma$ level of the dust emission ($1\sigma=23$\,\umjyb). The ordinate in the left is the intensity in units of Jy\,beam$^{-1}$ while the right ordinate indicates the corresponding dust brightness in K.}
    \label{fig:cont-assy}
\end{figure}

Figure \ref{fig:cont-assy} shows intensity profiles of the 1.3\,mm dust-continuum emission along the major and minor axes on both sides with respect to the disk center. 
The figure reveals overall a symmetric distribution of the dust continuum along the major axis of the disk, but with a slight depression of intensity ($<3\sigma$) at a radii of $0.18''$, deeper to the northwest direction than to the southeast direction. On the other hand, the profile along the minor axis presented in the lower panel of the figure shows an asymmetric distribution at $r>0.06''$. 

\begin{figure}[htbp!]
    \centering
     \includegraphics[width=0.8\linewidth]{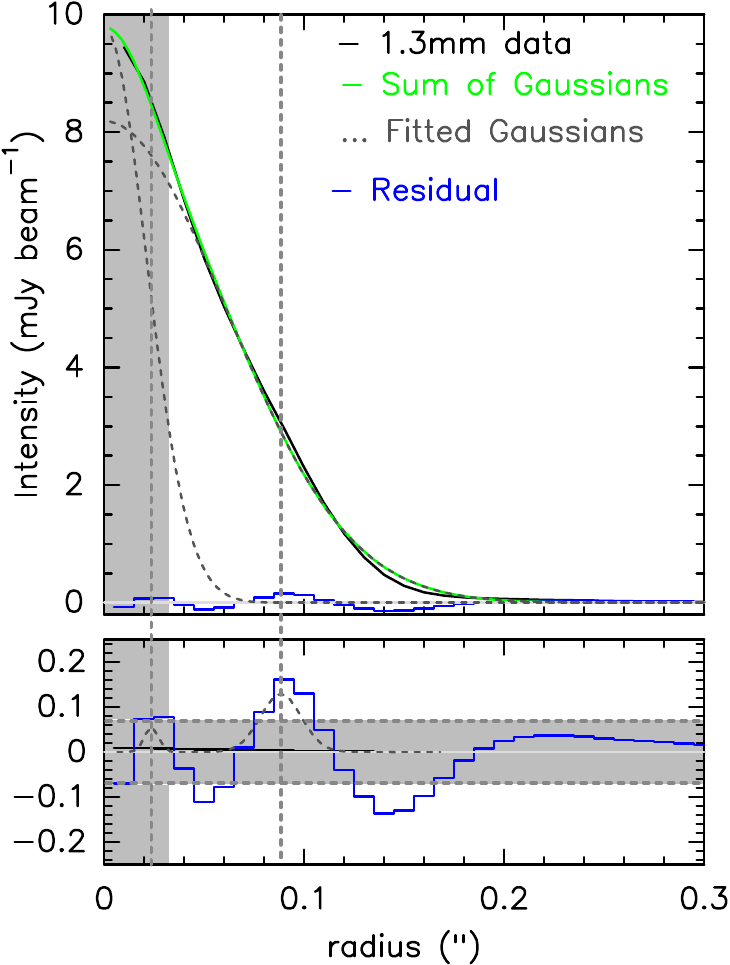}
    \caption{Radial profile of the 1.3\,mm dust intensity distribution azimuthally averaged in annuli of 0.01$''$ width in the plane of the disk. The data are shown in the black line, the two Gaussian fit components are drawn in black dashed lines and their sum is presented in the green line. The grey vertical band indicates the beam size. Possible ring position is marked with vertical dashed line.
The bottom panel enlarges the residual distribution along the radius which is also shown in the upper panel. The horizontal grey band marks the $\pm3\sigma$ level.}
    \label{fig:cont-radial-profile}
\end{figure}

Figure \ref{fig:cont-radial-profile} shows the radial intensity profile of the 1.3\,mm dust emission azimuthally averaged in annuli of 0.01$''$ ($\sim$ 1/4 of the beam) in the plane of the disk. The deprojection was performed using a  position angle of 121.5$^\circ$ and an inclination of 47$^\circ$, as derived from the 2D Gaussian fit presented above, assuming a flat-thin disk geometry. A single Gaussian function did not adequately fit the data. Therefore we performed a two Gaussian function fit to the radial profile. The results of the fit and its residual are presented in Figure \ref{fig:cont-radial-profile}. The residual profile shows two peaks and two valleys at different amplitudes which we refer to as ``ring'' and ``gap'', respectively.  To characterize the rings, we fit the residual with Gaussian functions. The inner ring has an amplitude at the level of 2.4\,$\sigma$, and the outer ring has an amplitude of $129\pm40\,\mu$Jy ($\sim6\pm2\,\sigma$) located at $\sim0.09''$ (14\,au) with a width of 10.3\,mas (1.6\,au), indicating that the amplitude level of the inner ring is not statistically significant. In addition, the width of the outer ring is only a fraction of the beam.  
Although the above analyses suggest the presence of possible ``ring'' and ``gap'' features in the protostellar disk, we note that this residual could be due to an azimuthally and radially local brightness enhancement, rather than a ring (i.e., an azimuthally uniform enhancement) since the intensity is averaged in the azimuthal direction. Additionally, because these features are close to our sensitivity level and resolution, we are unable to confirm them unambiguously.

\subsection{An analysis of Position-Velocity diagrams of \tco~ and \cho~}\label{sec:star_mass}

\begin{figure}
   \centering
    \includegraphics[width=\linewidth]{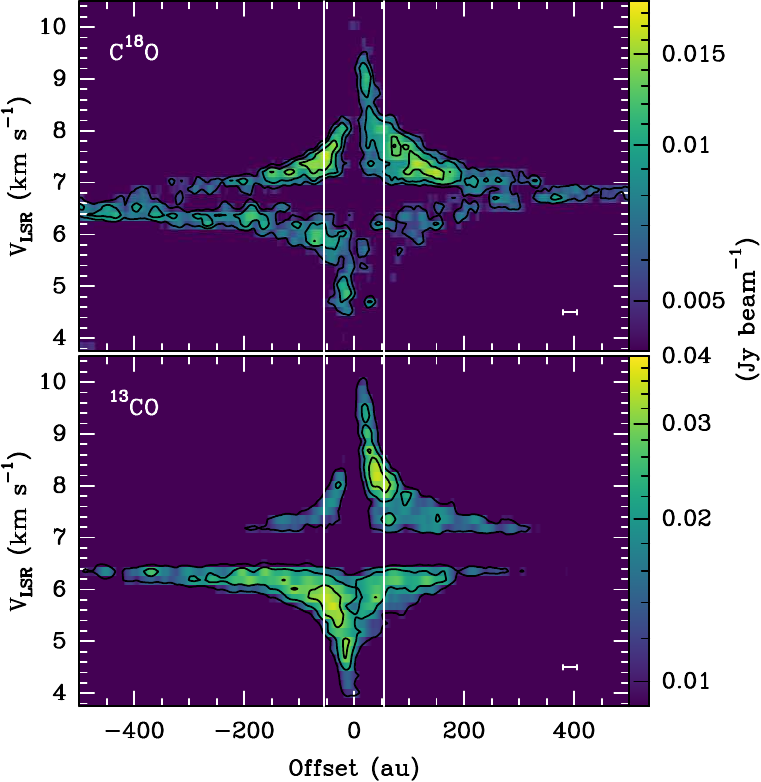}
    \caption{Position-Velocity diagram along the major axis of the dust disk for \tco~and \cho~emissions. 
    The contours start at 3$\sigma$, increasing in steps of $3\sigma$ with $1\sigma=1.8$\,\mjyb~for \cho~and $1\sigma=2.1$\,\mjyb~for \tco. The white bars on the lower right corner of each panel indicate the beam size. The vertical white lines mark the region where we search for the Keplerian motion.} \label{fig:c18o+13co_pv-full} 
\end{figure}

In this section, we attempt to estimate the protostellar mass by searching for a signature of Keplerian motion in the gas disk. For this purpose, we examine the position-velocity (PV) diagrams of the \tco~and \cho~emissions along the major axis of the dust disk (P.A.=122$^\circ$ east of north), within the width of one beam. As shown in Figure \ref{fig:c18o+13co_pv-full}, emissions are present in all 4 quadrants, indicating that the lines also trace other motions such as infall motions and/or outflows. 

We analyze these PV diagrams using the $pvanalysis$ module in the Python package Spectral Line Analysis/Modeling code\footnote{The Spectral Line Analysis/Modeling (SLAM) tool is publicly
available at https://github.com/jinshisai/SLAM.} \citep[SLAM,][]{Aso+Sai_2024}. The analysis and fitting processes are described in detail in the overview paper of the eDisk project \citep{Ohashi+edisk_2023}. Briefly, we first determine the emission edge and ridge positions (or velocities) from a one-dimensional cut of the PV diagram along the position (or velocity) axis. The ridge is determined from the peak derived by the 1D Gaussian fitting to the emission along the positional (or velocity) axis, while the edge is determined as the outermost contour of the emission above a given threshold. The derived positions and velocities $(r,v)$ can then be fitted with a power-law function: 
\begin{equation}\label{eq:pv}
\centering 
V_{rot}=|v-V_{sys}|=v_b(\frac{r}{r_b})^{-p}, 
\end{equation}
where $V_{sys}$ is the systemic velocity, $r_b$ is a characteristic radius, where $v_b$ is measured. 

\begin{table}
\centering
\caption{Best fit parameters to the PV diagrams for \tco~and \cho~emissions along the major axis of the dust disk.}\label{tab:PV-13co+c18o}
\centering 
\renewcommand{\arraystretch}{.8}
\setlength{\tabcolsep}{0.7em}
\begin{tabular}{ccllc}
\hline
& & \tco~ & \cho~& Unit\\
 \hline
 \multirow{3}{*}{Edge}  & $r_b$ & $31.2\pm1.0$ & $35.3\pm2.3$ & au \\
                        &  $v_b$ & $2.4\pm0.01$ & $2.3\pm0.01$ & \kms\\
                        &  $V_{sys}$& $7.0\pm0.02$ & $7.0\pm0.07$ & \kms\\
                        &  $p$& $0.66\pm0.13$ & $0.58\pm0.22$ \\
                        &  $M_\star$ & $0.38\pm0.01$ & $0.39\pm0.03$ & $M_\odot$ \\
\hline 
\multirow{3}{*}{Ridge} & $r_b$ & $16.1\pm1.1$& $13.0\pm2.3$ & au\\
                      & $v_b$ & $2.4\pm0.01$ & $2.3\pm0.01$ & \kms\\
                      & $V_{sys}$ & $6.8\pm0.15$& $6.8\pm0.11$ & \kms \\
                      & $p$ & $0.67\pm0.16$ & $0.49\pm0.22$\\
                      & $M_\star$ & $0.19\pm0.01$ & $0.15\pm0.03$ & $M_\odot$ \\
\hline 
\end{tabular}
\end{table} 

\begin{figure*} 
 \centering
 \includegraphics[width=0.77\linewidth]{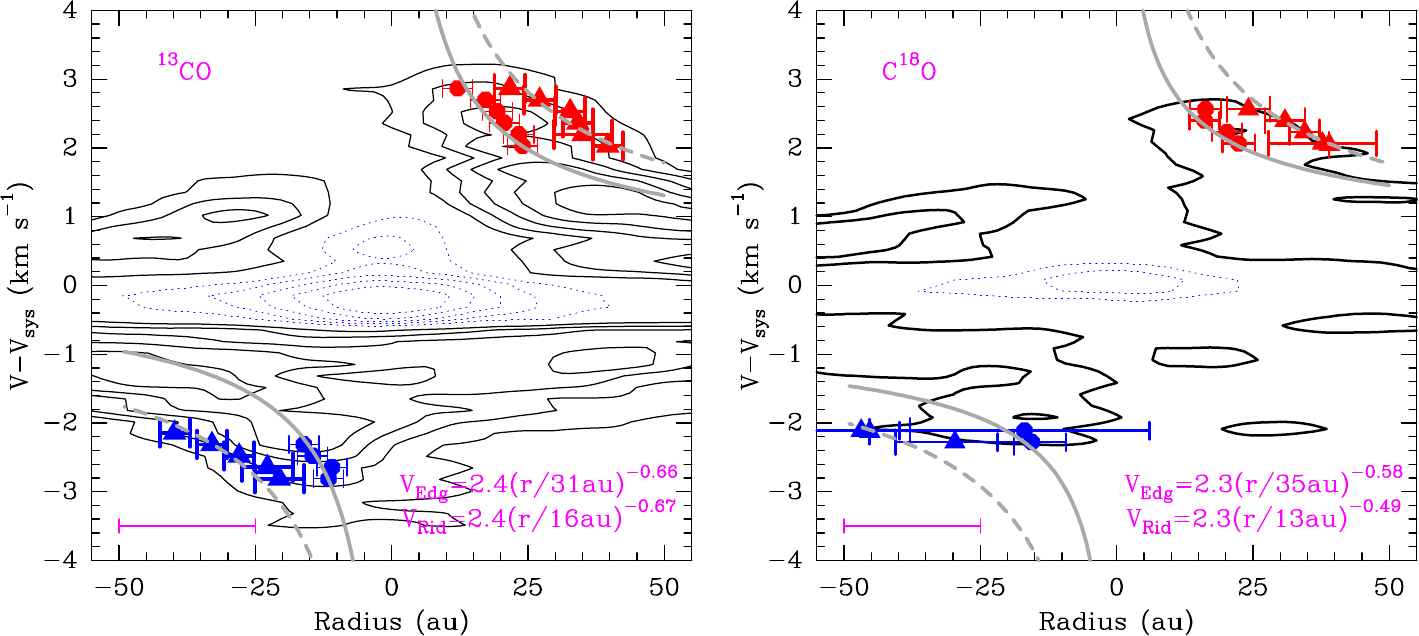}
 \caption{\textit{Left:} Position-Velocity diagram of $^{13}$CO (2--1) in the central region and the derived edge (triangle) and ridge (circle) points superimposed by their best-fit functions (solid and dashed lines, respectively). The magenta bar on the lower left corner shows the beam size, and the best-fit functions are indicated in the lower right corner (subscript ``Edg'' denotes the Edge and ``Rid'' denotes the Ridge). Contour levels are from $-15\sigma$ to $-3\sigma$ and $3\sigma$ to $15\sigma$ in steps of $3\sigma$, with $1\sigma$ is 2.1\,\mjyb. \textit{Right:} Same as the left panel but for \cho with contour levels are $-6\sigma$, -3$\sigma$, 3$\sigma$, and 6$\sigma$ ($1\sigma$ is 1.8\,\mjyb).}
    \label{fig:13co+c18o-pv-fit}
\end{figure*}


Figure \ref{fig:c18o+13co_pv-full} shows that the \tco~and \cho~PV diagrams consistently exhibit only red-shifted components in the first quadrant within the velocity range of \mbox{$v\geqslant8.5$\,\kms}. Conversely, in the third quadrant, only blue-shifted components are observed, specifically, in \tco~at v \mbox{$v\leqslant4.5$\,\kms} and in \cho~at \mbox{$v\leqslant4.7$\,\kms}.

We determine the edge and ridge data points at this velocity range and fit them with the power-law function (\ref{eq:pv}) to check whether the Keplerian disk is present in this small region. The ridge was derived using a threshold of $6\sigma$ for the case of \tco. For the case of lower signal-to-noise emission of \cho, we derived the ridge using a threshold of $3\sigma$. The best-fit parameters are $p=0.66\pm0.13$ and $p=0.67\pm0.16$ in the edge and ridge for the \tco~emission, respectively. In the case of \cho, these parameters are $p=0.58\pm0.20$, and $p=0.49\pm0.22$, respectively. The derived edge and ridge points and their corresponding best-fit function are presented Figures \ref{fig:13co+c18o-pv-fit} and the best-fit values are summarized in Tables \ref{tab:PV-13co+c18o}. The power-law fit indices agree with each other within their mutual uncertainties for the rotating gas motions to be closely consistent with Keplerian rotation ($V_{rot}\propto r^{-0.5}$) in the disk. 

Thus the stellar mass can be estimated under the assumption that the gases are in the Keplerian motion ($v=\sqrt{GM_\star/r}$) in the disk with inclination of $i=47^\circ$ around the central star. The stellar masses derived from the edge and ridge results of \tco~and \cho~ are consistent within their respective uncertainties. It is not clear which of the two techniques is more reliable. However, it is well known that the ridge fitting systematically underestimates, while the edge fitting overestimates the stellar mass \citep{Maret+Maury_2020, Aso+etal_2015}. Therefore, we have simply adopted the lowest and highest of derived values as the lower and upper limits of our stellar mass estimate as \mbox{$0.15\,\rm{M_\odot} < \rm{M_\star} < 0.39 \,M_\odot$}.

\begin{figure*}
\centering
\includegraphics[width=\linewidth]{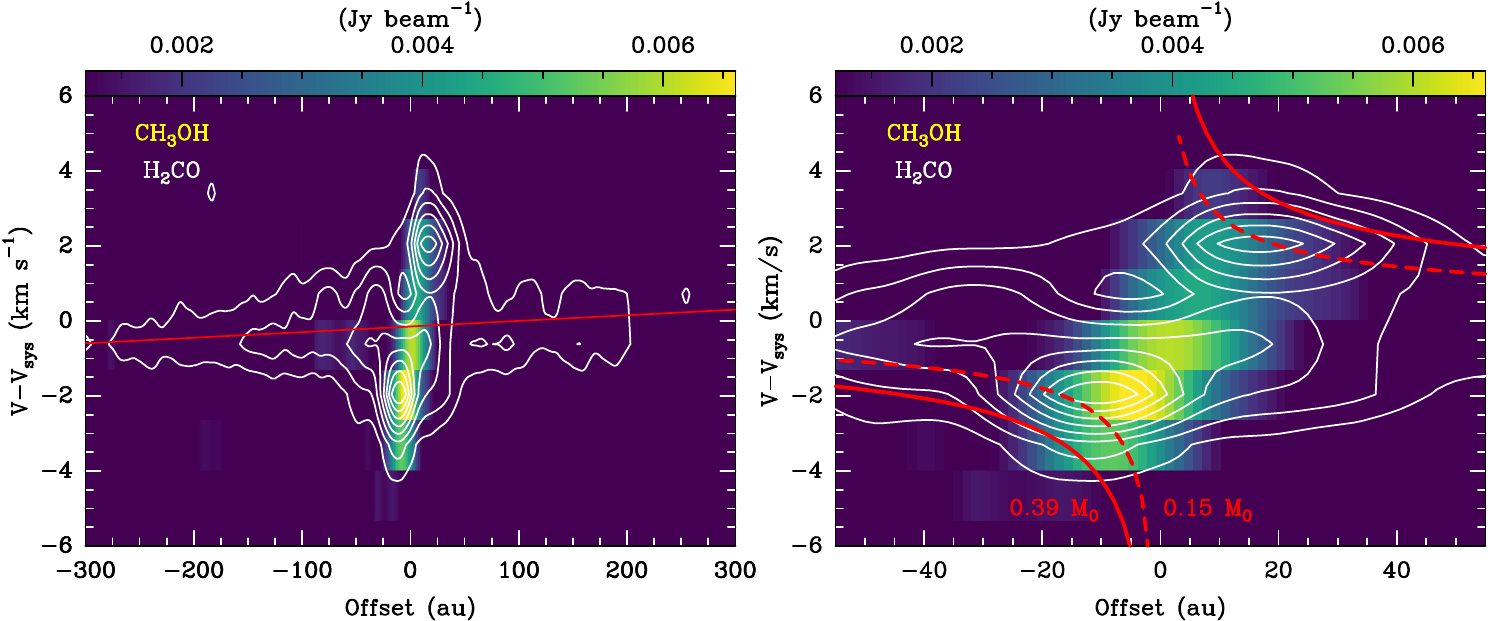}
\caption{\textit{Left:} Position-Velocity diagram of \metanol~(colors) and \hhco~(contours) cut along the major axis of the dust structure. The red line indicates the velocity gradient between the east and west directions. \textit{Right:} Same as the left panel but for the central region. The red curves show the Keplerian patterns of gas motion by a central protostar of 0.15\,M$_\odot$ and 0.39\,M$_\odot$. The contours start at 3$\sigma$ then increase in steps of 3$\sigma$.} \label{fig:ch3oh+h2co-pv}
\end{figure*}

As shown in Figure \ref{fig:h2co+ch3oh-moments}, the velocity gradient of \hhco~and \metanol~in the disk is well aligned with the major axis of the dust disk. Namely, these lines can be tracers of rotation motion in the disk. Figure \ref{fig:ch3oh+h2co-pv} presents PV diagrams of the \metanol~and \hhco~emission along the major axis of the dust disk in a width of 1 beam. On the large scale, the \hhco~emission appears in all quadrants in the PV diagram, suggestive of infalling motion, with a slight velocity gradient between the east and west, indicative of rotation. On the other hand, the \metanol~emission is mostly confined to the region of dust continuum emission from the disk and likely shows a linear velocity gradient. On the small scale, the \hhco~emission may show signatures of possible Keplerian motions in the disk. However, the poor velocity resolution of 1.34\,\kms~of \metanol~and \hhco~prevents us from performing a similar analysis to what we did for the \tco~and \cho~emission. Instead of making power law fitting of the data, we overlaid the Keplerian trajectories of gas motions by a central source of 0.15\,M$_\odot$ and 0.39\,M$_\odot$ in the PV diagrams for these tracers in the right panel of Figure \ref{fig:ch3oh+h2co-pv}, indicating that \hhco~ emission is consistent with the Keplerian motion.

\subsection{Jets and Outflows}\label{sec:jets_outflows}

\begin{figure*}
\flushright
\includegraphics[width=0.265\linewidth]{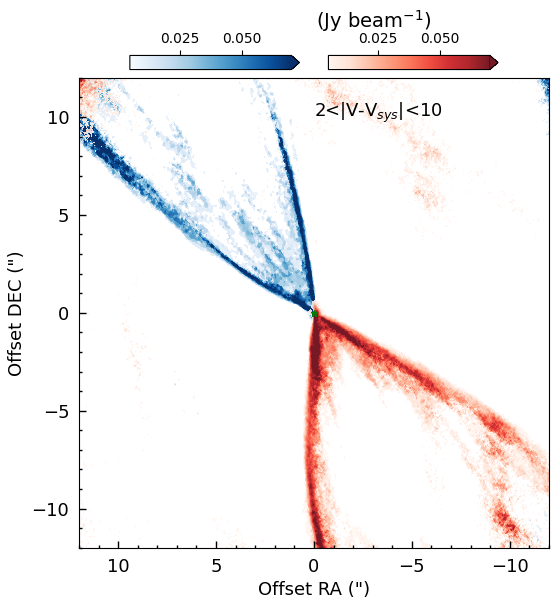}
\includegraphics[width=0.23\linewidth, trim =0cm -1.5cm 0cm 0cm]{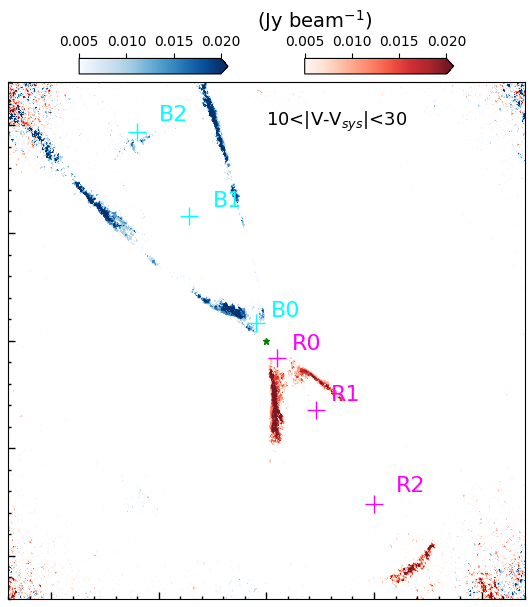}
\includegraphics[width=0.23\linewidth, trim =0cm -1.5cm 0cm 0cm]{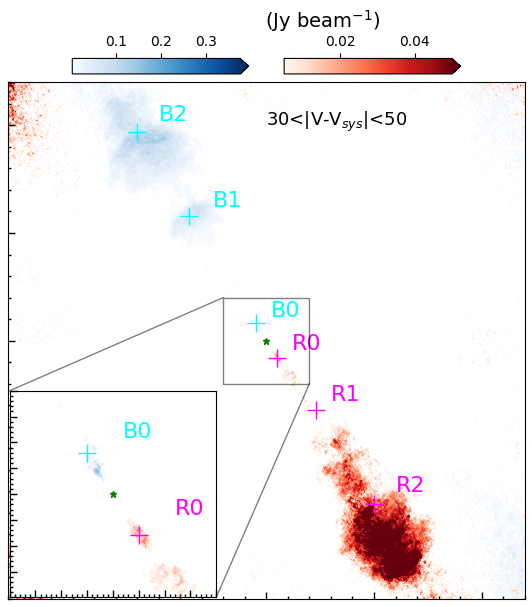}
\includegraphics[width=0.23\linewidth, trim =0cm -1.5cm 0cm 0cm]{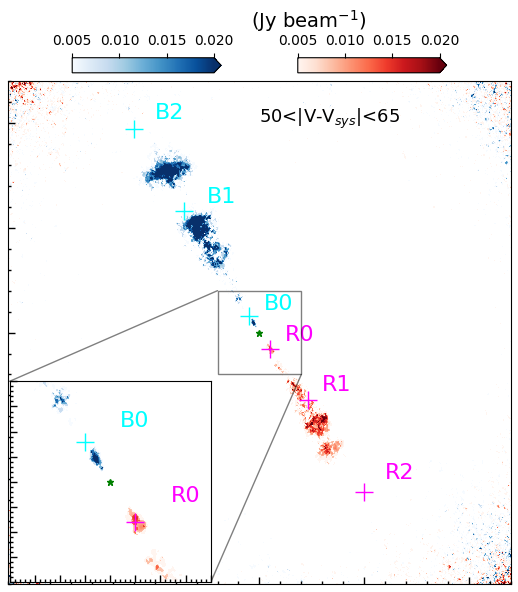}\\
\includegraphics[width=0.265\linewidth]{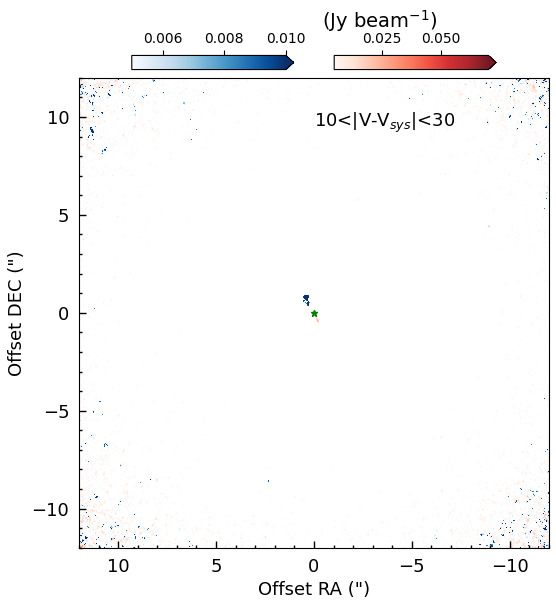} 
\includegraphics[width=0.23\linewidth, trim =0cm -1.5cm 0cm 0cm]{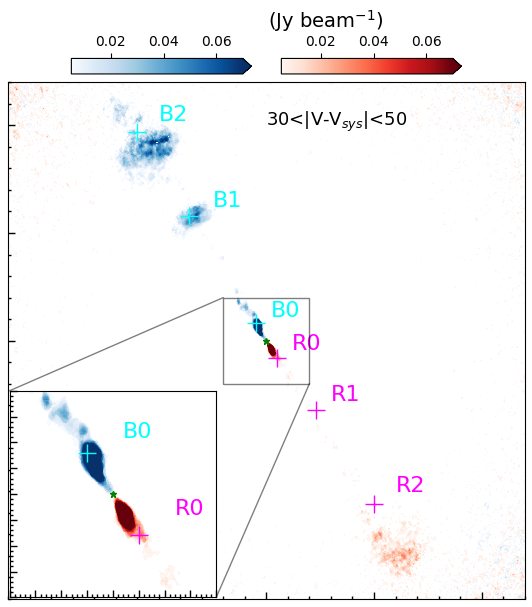}
\includegraphics[width=0.23\linewidth, trim =0cm -1.5cm 0cm 0cm]{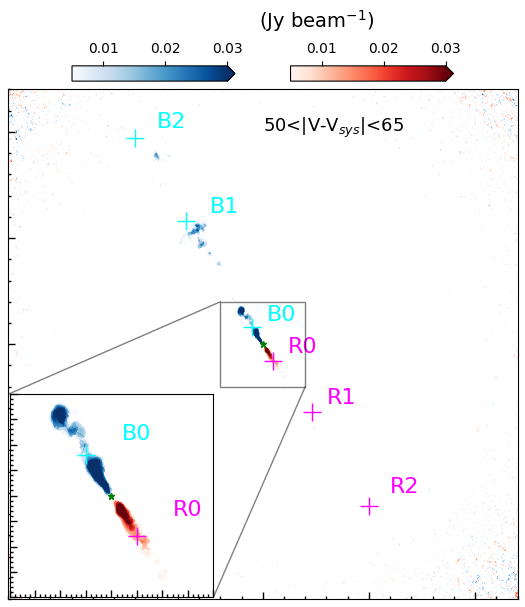}
\caption{Outflows and jets seen in CO and SiO emissions. The outflows and jets are shown in intensity maps integrated in the specific velocity range as indicated in each panel for \co~(\textit{upper panels}) and \sio~(\textit{lower panels}). The asterisk indicates the position of the central protostar. The central positions of knots (B0, B1, B2) and (R0, R1, R2) are marked with the cyan and magenta crosses in the blue-shifted and red-shifted emission, respectively.} \label{fig:12co+sio_regimes}
\end{figure*}

The molecular jets and outflows in IRAS 04166+2706 have been previously reported  in \co~and \mbox{SiO (2--1)} using the IRAM 30m telescope \citep{Tafalla+Santiago+etal_2004} and using the IRAM PdBI \citep{Santiago+etal_2009}. Those observations showed that the outflows and jets from IRAS 04166+2706 extend up to \mbox{$\sim 450''$}. Within $\sim120''$, the jets are found to include 7 pairs of knots at high-velocity of \mbox{$30<\mid V-V_{\rm{sys}}\mid<50$ \kms} \citep{Santiago+etal_2009}. \citet{Wang+Shang+etal_2014} then observed similar outflows and jets in the \mbox{CO(3--2)} line using ALMA. Recently, the outflows and jets in \co~and \sio~observations at 1.0$''$ resolution have been reported by \citet{Tafalla+Su+etal_2017}. Our observations also reveal the emission of \co~and \sio~of the two pairs of knots (R1, B1) and (R2, B2) within 24$''$ as reported by \citet{Wang+Shang+etal_2014}, and particularly resolve another pair of knots at the distance of $\sim0.5''-2''$ from the protostar (R0 and B0 marked in Figure \ref{fig:12co+sio_regimes}) thanks to the higher resolution of the present data. 

To look for more detailed structures in the outflows and jets, we present the moment 0 maps of the \co~(top) and \mbox{\sio~(bottom)} emissions integrated in the three velocity ranges of \mbox{$10<\mid V-V_{\rm{sys}}\mid<30$ \kms}, \mbox{$30<\mid V-V_{\rm{sys}} \mid<50$ \kms}, and \mbox{$50<\mid V-V_{\rm{sys}} \mid<65$} \kms~in Figure \ref{fig:12co+sio_regimes}. In the case of \co, we also present the intensity map of the low-velocity outflow component integrated in the range of \mbox{$2<\mid V-V_{\rm{sys}} \mid<10$ \kms~} in the leftmost panel of the figure. These intensity maps in specific velocity ranges are found to be very useful to reveal more prominent shapes of the conical bipolar outflows in the velocity range ($2<\mid V-V_{\rm{sys}} \mid<10$\,\kms) and the collimated jets in the range of $30<\mid V-V_{\rm{sys}} \mid<65$\,\kms. The integrated intensity map in the velocity range of \mbox{$10<\mid V-V_{\rm{sys}}\mid<30$ \kms} also reveals the presence of the faint emission of B2 knot and the outer bow of the R2 knot of the jet. At high-velocities, we clearly see the symmetric distribution of these two pairs of knots as reported earlier in \citet{Wang+Shang+etal_2014}. We also notice the presence of two bow shapes of the R2 knot in the \co~map in the velocity range of \mbox{$30<\mid V-V_{\rm{sys}}\mid<50$ \kms}. At the highest velocities, the narrower jets are found closer to the central source, separated from the B1 and R1 knots (marked as B0, and R0), much stronger in the \sio~than in the \co~emission. Particularly, the B0 and R0 knots seem reveal the wiggling motion as mentioned in the Section \ref{sec:obs_results}. 
These jets imply the presence of three episodic mass ejections (R0 to R2 and B0 to B2) with an average velocity of 40\,\kms~ within a radius of $11''$, yielding an average ejection period of

\begin{equation} 
\frac{11''\times 156 [\rm{pc}]}{\cos(i) \times 3\times 40\,[\rm{km/s}] / \sin(i)} \sim 73\, \rm{years},
\end{equation}
where $i=47^\circ$.

\begin{figure}
    \centering
     \hspace{0.5cm} (Jy beam$^{-1}$)\\
    \includegraphics[width=1.0\linewidth]{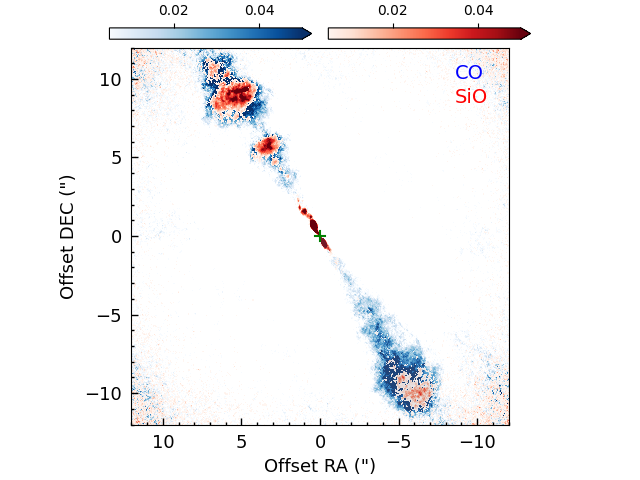}
    \caption{High velocity jets shown \co~(blue) and \sio~ emissions (red). }
    \label{fig:12co+sio-jets}
\end{figure}

Figure \ref{fig:12co+sio-jets} shows a comparison between the high-velocity components of \sio~and \co~emission in the range of \mbox{$30<|V-V_{\rm{sys}}|<65$ \kms}, indicating that the \sio~and \co~jets spatially overlap, and thus they might arise from the same mass ejection events. 

\begin{figure*}
    \centering
    \includegraphics[width=0.47\linewidth]{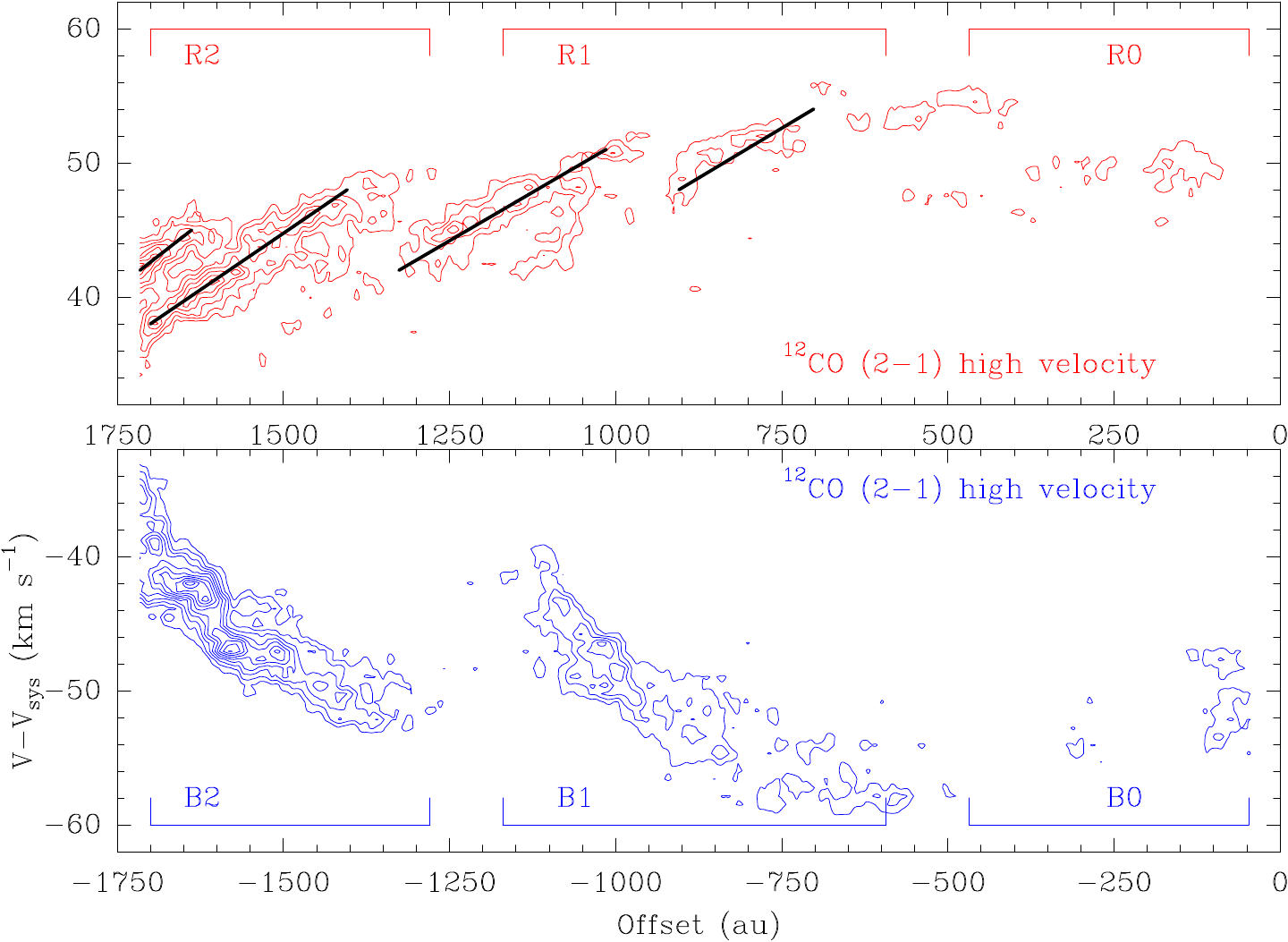}
     \includegraphics[width=0.47\linewidth]{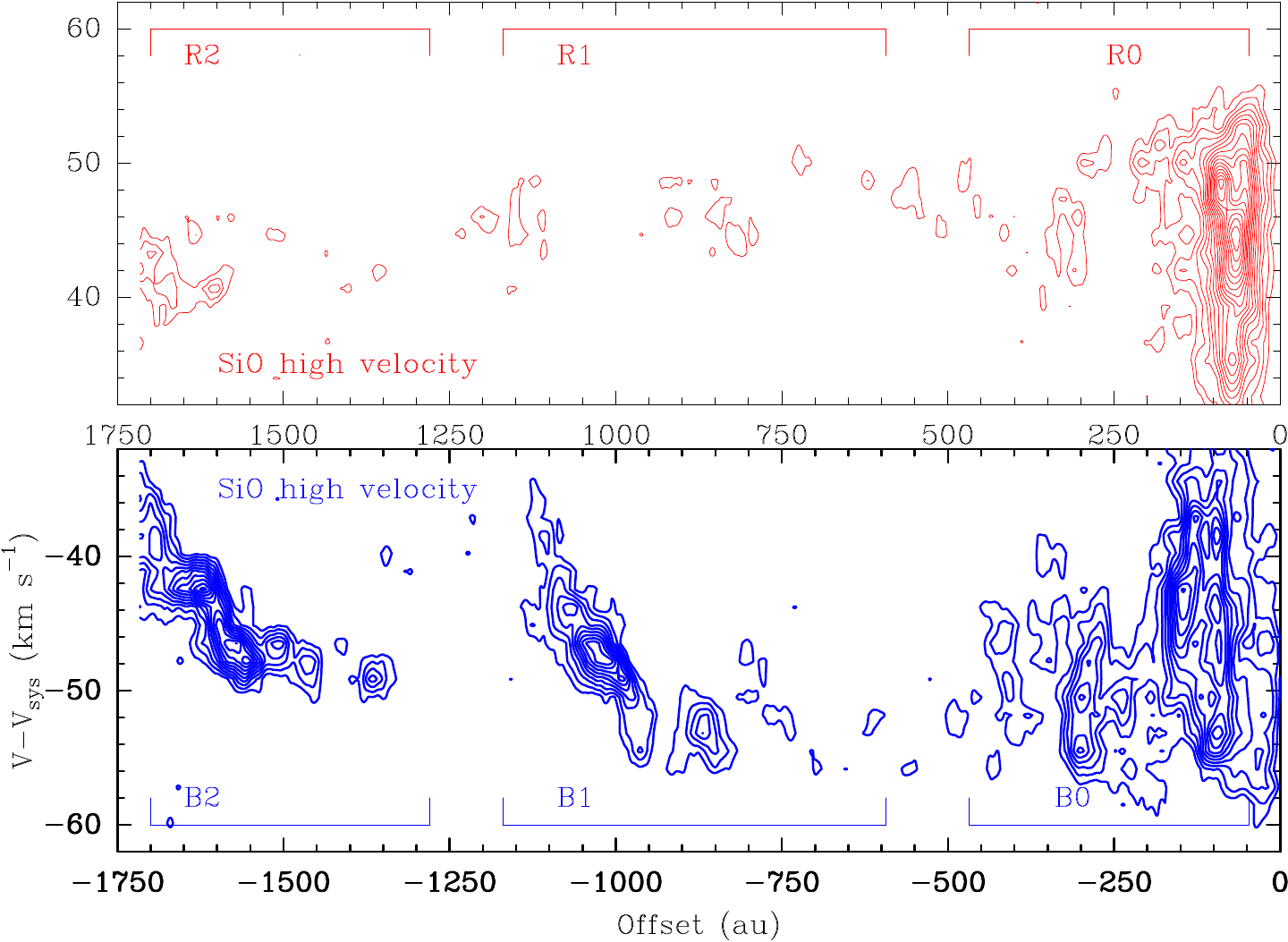}
    \caption{Position-Velocity diagrams of the high velocity jets for the \co~(left) and \sio~(right) emissions along the outflow axis (P.A=32$^\circ$). Contours are drawn from 3$\sigma$ with an increasing step of 2$\sigma$ where 1$\sigma$ is 1.1\,\mjyb~ for \sio~ emission and 2.0\,\mjyb~ for \co~ emission.}
    \label{fig:12co+sio-pv-EHV}
\end{figure*}

Figure \ref{fig:12co+sio-pv-EHV} shows the PV diagrams of the high-velocity jets of \co~and \sio~emission along the jets/outflows axis (P.A.=32$^\circ$). At the high velocities, both \co~and \sio~show a saw-tooth feature exhibiting episodic ``slow head'' and ``fast tail'' properties - the ``head'' of the mass ejection, which is further in the distance than the ``tail" from the central star, moves slower than the ``tail", as already reported by \citet{Santiago+etal_2009}. The difference in the velocity between the ``head'' and the ``tail'' is about 15--20 \,\kms. The features were interpreted as arising from gas ejected sideways in a series of internal shocks inside the pulsating jets \citep{Tafalla+Su+etal_2017} or by a spherically expanding wind with axial density concentration \citep{Wang+Shang+etal_2019}. 
Previous observations by \citet{Wang+Shang+etal_2014} identified the two pairs of knot (B1, R1) and (B2, R2), but not as clearly as our observations with $\sim 40$ times better spatial resolution. Especially, in the red-shifted part of the \co~emission, each knot is found to contain two features that were marked with black lines in the upper left panel of Figure \ref{fig:12co+sio-pv-EHV}. 
We also note that there are two other knots (B0, R0) close to the central protostar, each likely contains also two continuous ejection events at distances of about 100\,au and 300\,au (see Figure \ref{fig:12co+sio-pv-EHV}), respectively. Based on the B0 and R0 knots, we estimate that the most recent ejection events may have occurred only about 20–25 years ago.

\begin{figure*}
    \centering
     \includegraphics[width=\linewidth]{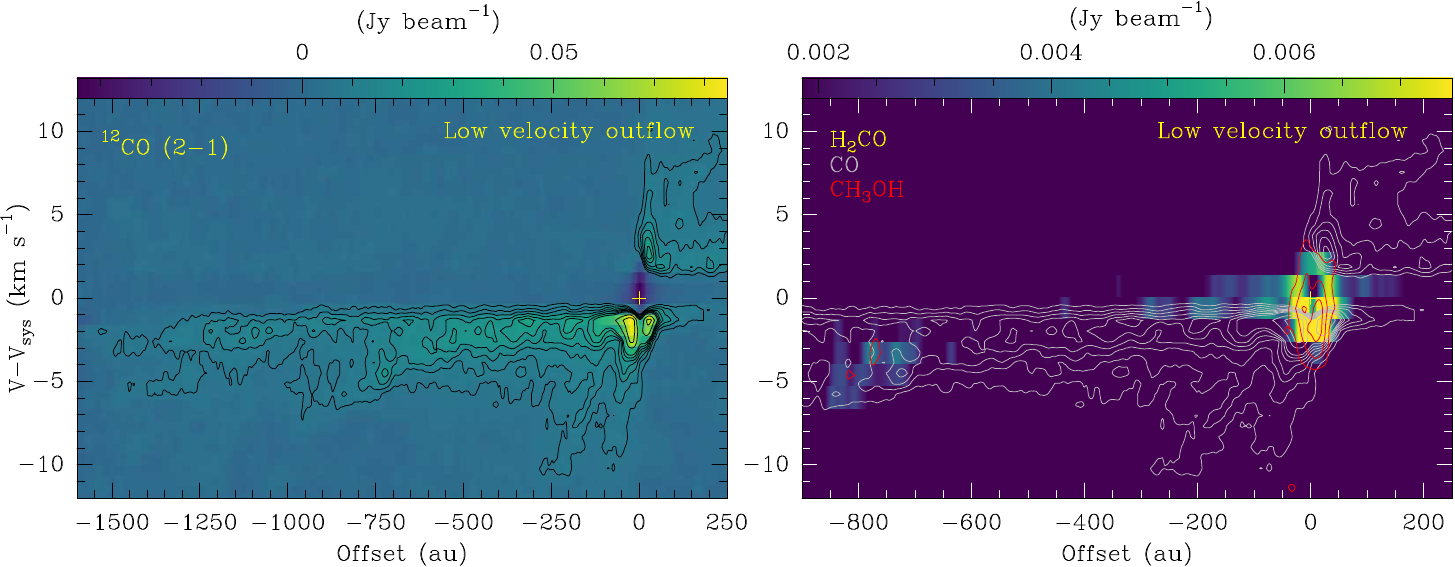}
    \caption{\textit{Left:} Position-Velocity diagram of low-velocity outflows as seen in the \co~ emission along the outflow axis. The contours level starts at 3$\sigma$ and increases by a step of 3$\sigma$ to 15$\sigma$ and then a step of 9$\sigma$ (1$\sigma$=2.0\,\mjyb). The red cross marks the position of the central protostar. \textit{Right:} Position-Velocity diagrams along the outflow axis of \hhco~(colors) and \metanol~(red contours), superimposed by the \co~ emission (grey contours). The contours start at 3$\sigma$ and then increase by a step of 3$\sigma$ (1$\sigma$ is 0.6\,\mjyb) in the case of \metanol. The contour levels of \co~are the same as in the left pannel.}
    \label{fig:pv_SHV}
\end{figure*}

The PV diagram of the \co~ low-velocity outflow component (left panel of Figure \ref{fig:pv_SHV}) shows that the mean velocity of the outflow is within 5\,\kms~ with respect to the systemic velocity. The blue-shifted emission is much stronger than the red-shifted emission. Figure \ref{fig:pv_SHV} (right panel) shows the PV diagrams of \hhco~and \metanol~along jets/outflows axis superimposed by the CO outflow emission, suggesting that the \hhco~and \metanol~features in the velocity range of  \mbox{$V_{\rm{LSR}}\sim 0-4$\,\kms} at large distance from the central protostar ($\sim 800$\,au)  can be likely outflow-related.

\section{Discussion} \label{sec:dis}
\subsection{Brightness asymmetry, dust mass in the disk, and disk stability}
In Section \ref{sec:dust-morp}, we have shown that the 1.3\, mm observation reveals a disk structure with a radius of $\sim22$\,au, with the presence of a brightness asymmetry along the minor axis of the disk. 
This asymmetric emission distribution along the minor axis has been also seen in other protostellar disks which are on close edge-on geometry, such as HH212 mms \citep{Lee-CF+etal_2017} and the eDisk targets (CB 68; \citet{Kido+eDisk_2023}, L1527IRS; \citet{vanHoff+eDisk_2023}, IRAS 04302+2247; \citet{Lin+eDisk_2023}, and GSS30 IRS3; \citet{Santamaria-Miranda+eDisk_2023}). Recent 3D radiative transfer (RADMC-3D) calculations by \citet{Takakuwa+eDisk_2024} demonstrated that this asymmetric distribution of dust disk emission can be caused by the combined effect of vertically extended dust emission above the mid-plane, a so called flared distribution of the dust and high optical depth of the dust emission. Therefore it is thought that for IRAS 04166+2706, its inclined disk is probably flared in a way that its far and near sides are on the northeastern (NE) and southwestern (SW) parts of the disk, respectively, as expected from the outflow patterns blueshifted to the northeast and redshifted to the southwest. This geometry of the disk would result in its northeastern part being brighter and the southwestern part being fainter, as our observing line-of-sight would preferentially observe the hotter part of the northern disk and cooler part of the southeastern disk. We measured an integrated flux ratio of 1.17 between the northeastern (NE) and southwestern (SW) hemispheres, indicating that the far-side hemisphere is 17\% brighter than the near side. This is broadly consistent with the asymmetry seen in the synthetic continuum image from \citet{Takakuwa+eDisk_2024}. Their models suggest that, for a disk with $M_{\rm{disk}} \leq 0.14 M_\star$, a flaring index of $h/r \propto r^{0.3}$ is required to produce a clearly asymmetric intensity profile along the minor axis. An even higher flaring index is needed for disks with lower mass, such as in the case of IRAS 04166+2706. However, other reasons, such as non-axisymmetric internal structures, cannot be ruled out. It is possible that the large-scale spirals as shown in the Figure \ref{fig:cont}(b) contribute to the observed asymmetric distribution. A more quantitative comparison between the observed asymmetry and the radiative transfer model is needed to determine the origin of the asymmetry in the disk.

On the other hand, our estimated dust mass in the disk of the order of 25 to 50 M$_\oplus$ is found to be useful to make a discussion on the potential for planet formation in the disk. As discussed by \citet{Tychoniec+etal_2020}, the planet formation efficiency is of the order of $15\%$ assuming that the planet starts to form in the Class 0 disk. In this scenario, the disk mass of IRAS 04166 +2706 is sufficient to form a core (of a planet)  with a mass of 4 - 7 M$_\oplus$ in the disk. Moreover, this young disk at the Class 0/I stage is still accreting the material from its natal envelope. Then the amount of the material both in the solid and gas in the disk may be replenished, thus increasing the material available for planet formation to form a giant planet core in the future. In this case, the solid mass in the disk of IRAS 04166+2706 is likely sufficient to form a giant planet core of $\geqslant7_\oplus$.

The stability of a disk against gravitational collapse can be examined by using Toomre's criterion of \citep{Toomre_1994}
\begin{equation} 
Q=\frac{c_s\Omega}{\pi G \Sigma} = 2 \frac{M_\star}{M_{\rm{disk}}}\frac{H}{R}, 
\end{equation}
where $M_\star$ is the mass of the central star, $M_{\rm{disk}}$ is the mass of the disk, R is the radius of the disk, and $H=c_s/\Omega$ while $c_s$ is sound speed and $\Omega=\sqrt{G M_\star/R^3}$. A geometrically thin disk would be gravitationally unstable if its Toomre number of $Q<1-1.5$ depending on the type of disturbances in disk \citep{Durisen+etal_2007}. 
In the case of IRAS 04166+2706, adopting R=22\,au  $M_\star$=0.15--0.39\,$M_\odot$, $M_{\rm{disk}}$=0.015 $M_\odot$ and 0.008 $M_\odot$, and the sound speeds $c_s$ for the dust temperature of 20\,K and 34\,K, we derive the Toomre number of $\sim1.4$ to 5.6. This result indicates that the disk of IRAS 04166+2706 is stable, or at most, marginally unstable but we emphasize that there are still significant uncertainties due to the assumptions about optical depth, dust opacity, and dust temperature.

\subsection{Wiggling motion in the jets and knotty jets}
In this section, we discuss other indirect signatures for substructures in the disk. 
The wiggling motion in the high-velocity jet component seen in the \sio~emission as shown in the right panel of Figure \ref{fig:sio-moments} is one of them. The wiggling structure observed in the molecular outflows/jets is not fully understood in the field. It can be explained in terms of the variability of mass ejection directions \citep[e.g.][]{Gueth+etal_1996, Podio+etal_2016} caused by orbital motion of a companion \citep{Masciadri+Raga_2002}, or tidal interactions between the protostellar disk and noncoplanar binary component \citep{Terquem+etal_1999},
which all require a companion (binary system). \citet{Frank+Ray_2014} also proposed a mechanism that such mass ejection angle variability can be caused by the ejection mechanism itself (e.g. controlled by the disk B-field) in a single star system. At the very high angular resolution observations of 8\,au in our eDisk observations, we could not resolve any binary component in the IRAS 04166+2706 system, thus the former hypothesis may not be plausible to explain the wiggling jets observed in the \sio~emission. However, it may be interesting to examine in the future whether the presence of an unresolved companion, like planets in the system, can affect any pattern of jet gas motions from a protostellar object.

The knotty jets observed in \co~and \sio~may indicate episodic mass ejections of a period of \mbox{$\sim73$} years \citep[two times smaller than the value of 150\,years estimated by][from the larger scale observations which resolved 7 knots within a radius of $\sim 60''$]{Tafalla+Su+etal_2017}. Our observations also revealed the presence of a pair of recent mass ejection events occurred $\sim 20-25$ years ago.  
We note that the removal of angular momentum and disk mass via magnetocentrifual wind can lead to the formation of local substructures such as rings or gaps \citep{Pascucci+etal_2022}.
Future identification of substructures in the disk would be highly useful to examine a possible link between their formation in the disk and the knotty jets in the IRAS 04166+2706 in relation with an episodic angular momentum removal.  

\section{Summary}\label{sec:sum}
This paper presents the results of ALMA observations of 1.3\,mm dust continuum at an angular resolution of $0.05''$, and molecular lines (\co, \tco, \cho, \hhco~and \metanol) at an angular resolution of $\sim 0.15''$ towards the Class 0 source IRAS 04166+2706. Our main results are summarized as follows: 

\begin{itemize}
\item The observation at 1.3\,mm revealed a dust disk structure of a radius of 0.15$''$ ($\sim$22\,au), its major axis at P.A.=122$^\circ$,  and an inclination of 47$^\circ$. The mass of the disk is estimated to be  \mbox{$M_{\rm{disk}}\sim1.5\times10^{-2}\,\rm{M}_\odot$}. 

\item The dust disk shows an asymmetric brightness distribution along the minor axis in the sense of the brightness peaking to the northeast of the disk geometric center. We are cautious to interpret this as an existence of a possible substructure in the disk, as such brightness asymmetry can be easily caused by the combined effect of a flared distribution of the dust and high optical depth of the dust emission.

\item Multiple molecular lines such as the \tco, \cho, \hhco~and \metanol~lines are found to probe the rotating gas motions in the disk. Particularly \cho~line likely traces the Keplerian motion, enabling to estimate the mass of the protostar mass to be $0.15\,\rm{M_\odot} < \rm{M_\star} < 0.39 \,M_\odot$.

\item Molecular lines of \co~ and \sio~ show energetic, knotty jets as already found in previous studies. However, our results showed the innermost jet/outflow knots, B0 and R0, close to the central protostar, which may have very recently occurred only 20–25 years ago. 

\end{itemize}

\section*{Acknowledgments}
We are grateful to the anonymous reviewer for valuable comments. This paper makes use of the following ALMA data: ADS/JAO.ALMA\#2019.1.00261.L. ALMA is a partnership of ESO (representing its member states),
NSF (USA) and NINS (Japan), together with NRC (Canada), MOST and ASIAA (Taiwan), and
KASI (Republic of Korea), in cooperation with the Republic of Chile. The Joint ALMA Observatory
is operated by ESO, AUI/NRAO and NAOJ. The National Radio Astronomy Observatory is
a facility of the National Science Foundation operated under cooperative agreement by Associated
Universities, Inc.
C.W.L. was supported by the Basic Science Research Program through the National Research Foundation of Korea (NRF) funded by the Ministry of Education, Science and Technology (NRF-2019R1A2C1010851), 
and by the Korea Astronomy and Space Science Institute grant funded by the Korea government (MSIT; project No. 2024-1-841-00).
J.J.T. acknowledges support from NASA XRP 80NSSC22K1159. Z.Y.L is supported in part by NASA 80NSSC20K0533 and NSF AST-2307199 and AST-1910106. J.K.J. and S.G. acknowledge support from the Independent Research Fund Denmark (grant No. 0135-00123B). S.T. is supported by JSPS KAKENHI Grant Numbers 21H00048 and 21H04495. This research is funded by the Vietnam Academy of Science and Technology under grant number VAST08.02/25-26. This work was supported by NAOJ ALMA Scientific Research Grant Code 2022-20A. This work was partly supported by a grant from the Simons Foundation to IFIRSE, ICISE(916424;N.H.). L.W.L acknowledges support from NSF AST-2108794. ZYDL acknowledges support from NASA 80NSSCK1095, the Jefferson Scholars Foundation, the NRAO ALMA Student Observing Support (SOS) SOSPA8-003, the Achievements Rewards for College Scientists (ARCS) Foundation Washington Chapter, the Virginia Space Grant Consortium (VSGC), and UVA research computing (RIVANNA).
N.O. and C.F. acknowledge support from National Science and Technology Council (NSTC) in Taiwan through NSTC 113-2112-M-001-037 and from Academia Sinica Investigator Project Grant AS-IV-114-M02. 

\vspace{5mm}

\bibliography{iras04166_firstlook}{}
\bibliographystyle{aasjournalv7}
\appendix
\section{Channel maps of main molecular lines}\label{sec:channel-maps}
\counterwithin{figure}{section}
\setcounter{figure}{0}

\begin{figure*}[ht!]
\centering
  \includegraphics[width=0.84\linewidth]{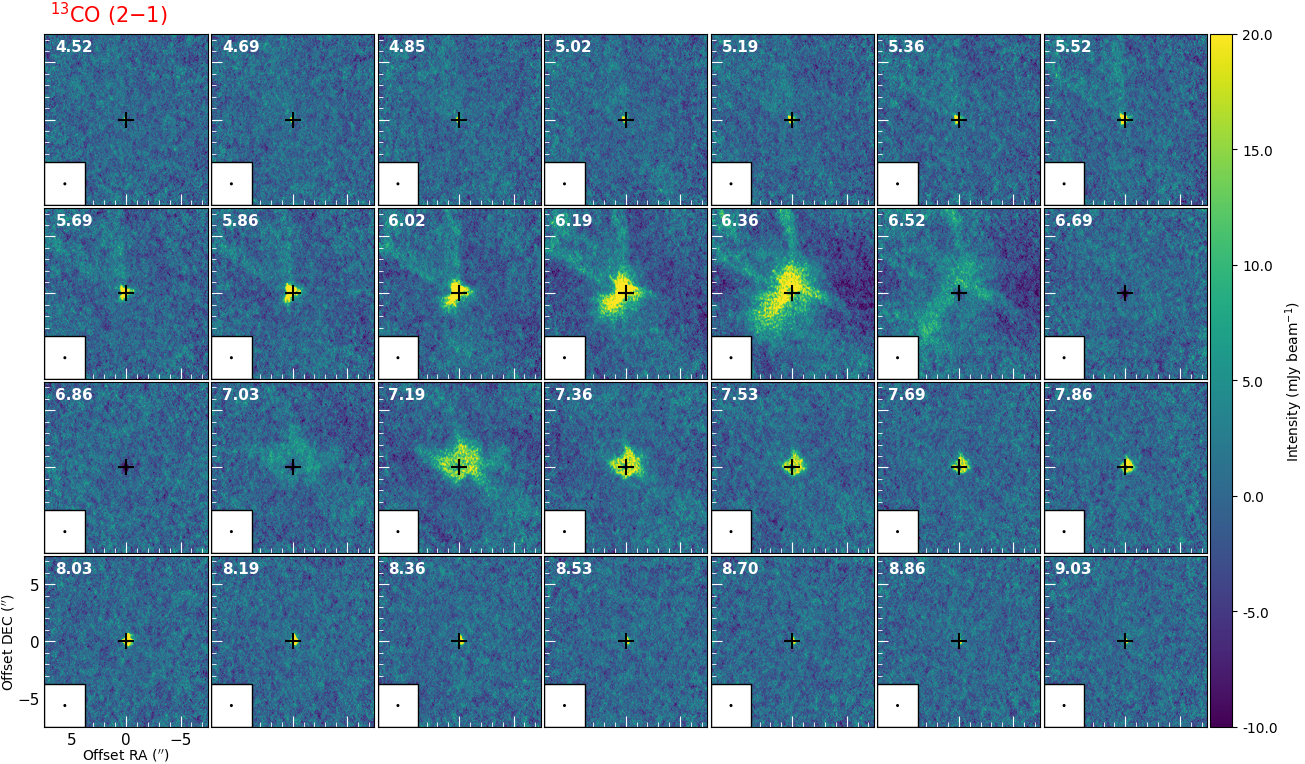}
  \includegraphics[width=0.84\linewidth]{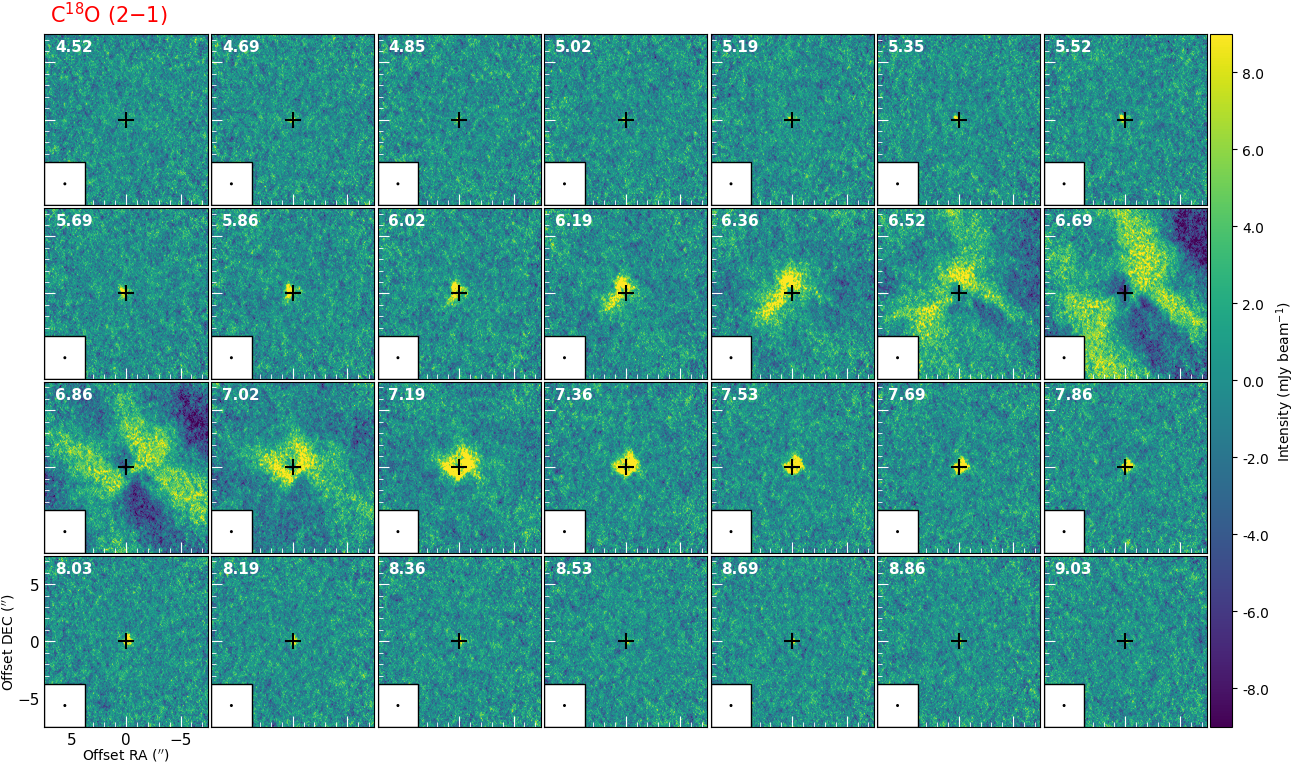}
  \caption{Channel maps of \tco~ (upper) and \cho~ (lower) lines. 
  The number inserted in the upper left corner is velocity in \kms.}\label{fig:channel-13co+c18o}
 \end{figure*}

\begin{figure*} 
\centering
\includegraphics[width=1\linewidth]{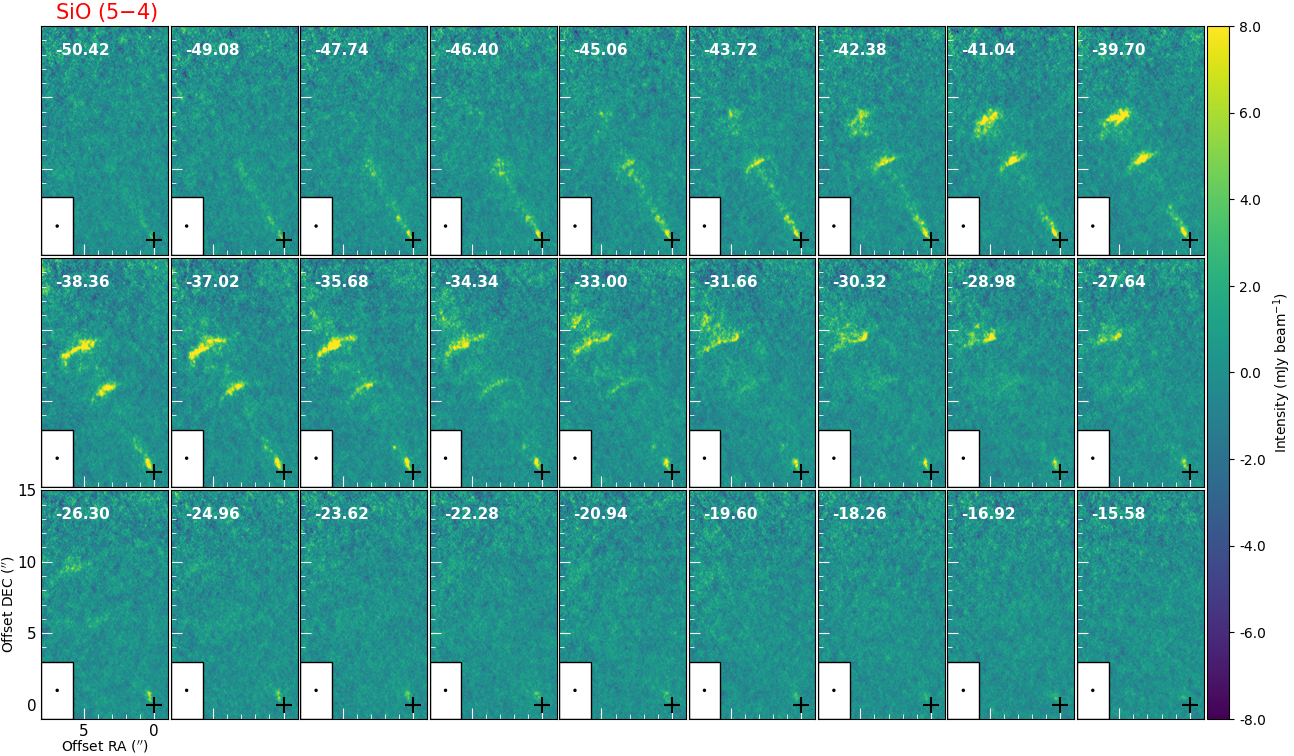}\\
\includegraphics[width=1\linewidth]{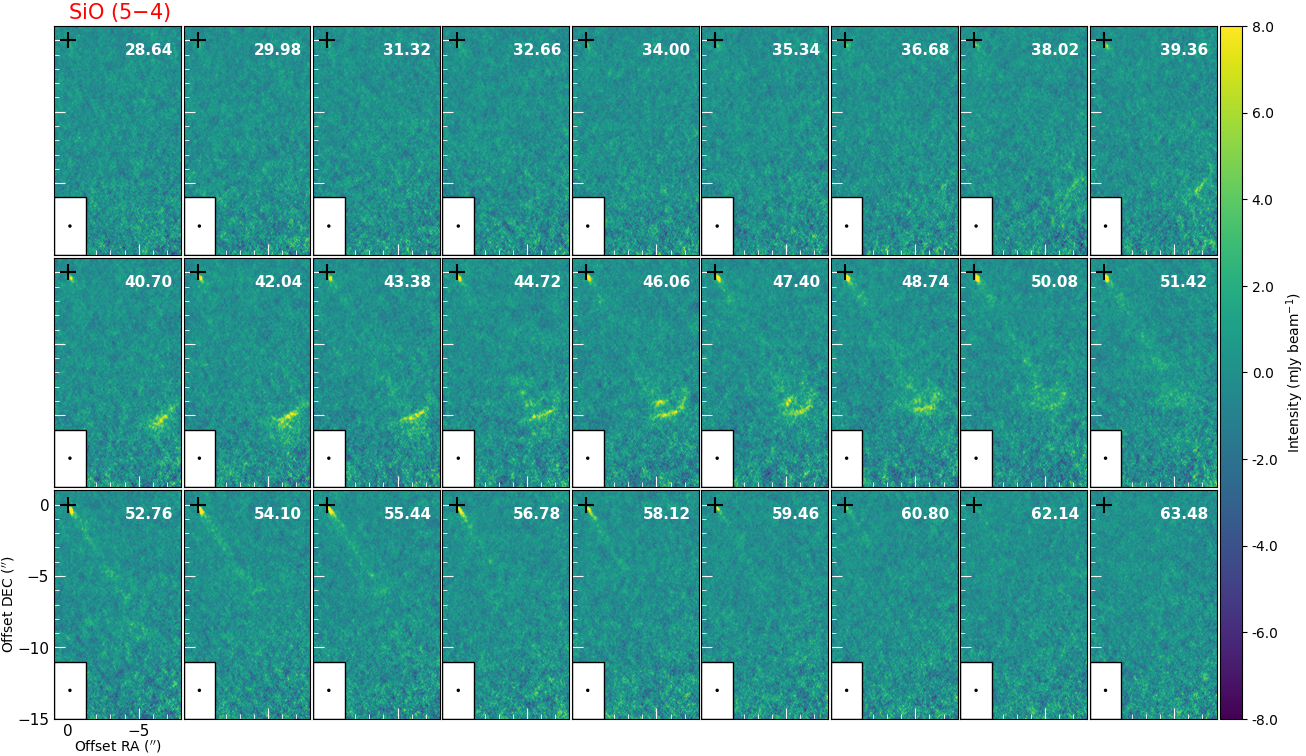}
\caption{Channel maps of \sio~emission in the range of $20<\mid V-V_{sys} \mid<60$\,km\,s$^{-1}$ in the blue-shifted (upper) and red-shifted (lower) sides.} \label{fig:channel-sio}
\end{figure*}

\begin{figure*} 
\centering
    \includegraphics[width=1.0\linewidth]{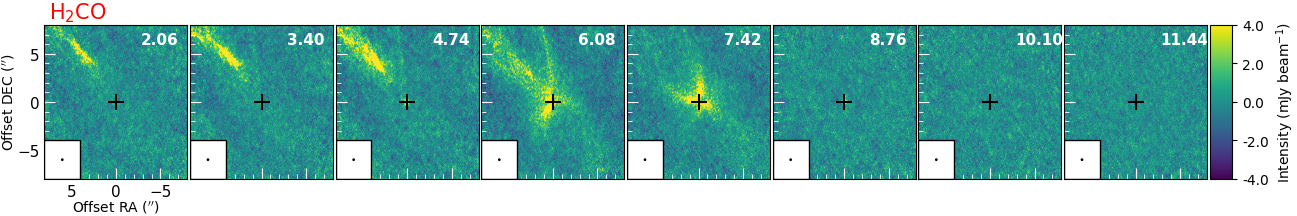}
    \includegraphics[width=1.0\linewidth]{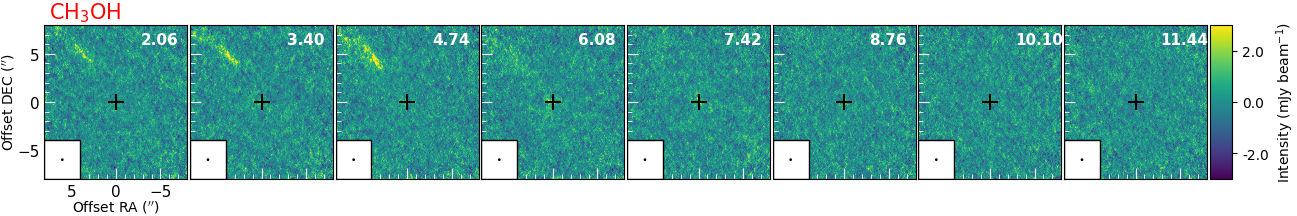}
    \includegraphics[width=1.0\linewidth]{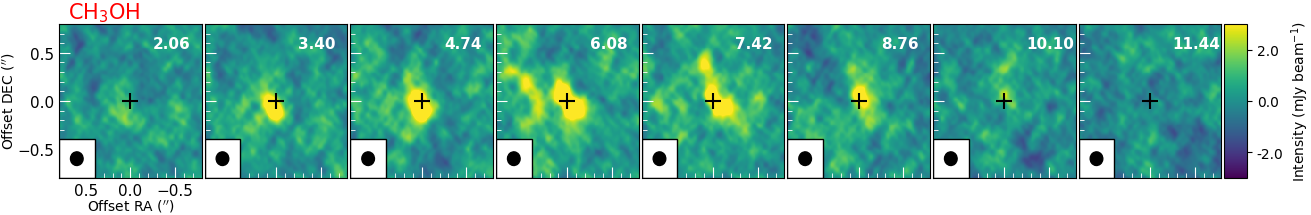}
    \caption{Channel maps of \hhco~ and \metanol~ emissions. The bottom-most panels show a closer view of the center region of \metanol.}\label{channel-h2co+ch3oh}
\end{figure*}

In this section, we present the channel maps of the main molecular lines that we presented and discussed in Section \ref{sec:obs_results}. Channel maps of \tco and \cho~are presented in Figure \ref{fig:channel-13co+c18o}. Channel maps of \sio~are presented in Figure \ref{fig:channel-sio}, and channel maps of \hhco and \metanol~are presented in Figure \ref{channel-h2co+ch3oh}.

\clearpage
\section{Other molecular lines}\label{sec:other_lines}
\counterwithin{figure}{section}
\setcounter{figure}{0}
\counterwithin{table}{section}
\setcounter{table}{0}
In this section, we present all the other molecular lines that we observed in the eDisk program towards IRAS04166+2706. They are \hhhco, \hhhco, \so, \Cyclopropenyo, \Cyclopropenyt, \Cyclopropenytt, and \dcn. All images are produced using $robust$ parameters of 2.0. We summarize the main parameters of these observations in Table \ref{tab:other_lines} and present their moment maps in Figures \ref{fig:c3h2+dcn} and \ref{fig:h2co+so}. The moment 0 and moment 1 maps of these 7 lines have been computed using a threshold of 3$\sigma$ in the velocity range of \mbox{$V_{\rm{LSR}}$=3--11 \kms}. 

Figure \ref{fig:c3h2+dcn} shows the moment 0 and moment 1 maps of three c-C$_3$H$_2$ lines and DCN emission. The \Cyclopropenyo~and \Cyclopropenyt~lines are well detected and show X-shape like features in the center region of $\sim 3.0''$ as in the case of \hhco. Also, the moment 1 maps reveal the complicated kinematics mixed between the red- and blue-shifted emission at the same offset position. The DCN emission is detected in the central region of $\sim 1.25''$ with complex kinematics. 

As shown in Figure \ref{fig:h2co+so}, the two lines of \hhhco~and \hhhhco~(upper-level energy $E_u$=68\,K) are well detected in the central region close to the protostar but the extent is much narrower than that of its lower energy level line of \hhco ($E_u$=21\,K). The \so~line is also well detected in the very compact center region and peaks outside the central protostar at an offset of $\sim 0.1''$ in the southwest quadrant. In addition, the blueshifted outflow-related feature has been detected in all three lines like in the case of \metanol~(and lower energy level transition of H$_2$CO). This likely supports that the emission is caused by shocks from the outflow. 

\begin{table}[h!]
\centering
\caption{Summary of ALMA observations of the SO, DCN, H$_2$CO and c-C$_3$H$_2$ molecular lines}\label{tab:other_lines}
\centering 
\renewcommand{\arraystretch}{1.1}
\setlength{\tabcolsep}{1.em}
\begin{tabular}{cccccc}
\hline\hline
Lines & Frequency & Velocity resolution  & E$_u$  & Beam (P.A) & Noise  \\ 
    & (GHz) & (\kms) & (K) &  & (mJy beam$^{-1}$) \\
\hline
\hhhco & 218.760056 & 0.167 & 68.1 & $0.17''\times0.14''(\rm{7^\circ})$ & 1.97 \\
\hhhhco & 218.475642 & 1.34 & 68.1 & $0.17''\times0.14''(\rm{7^\circ})$ & 0.71 \\  
\Cyclopropenyo & 217.829416 & 1.34 & 38.6& $0.17''\times0.14''(\rm{6^\circ})$ & 0.77\\
\Cyclopropenyt & 217.947320 & 1.34 & 35.4 & $0.17''\times0.14''(\rm{6^\circ})$ & 0.71\\ 
\Cyclopropenytt & 218.167717& 1.34 & 35.4 & $0.17''\times0.14''(\rm{6^\circ})$ & 0.71\\ 
\so & 219.953477 & 1.34 & 35.0 & $0.18''\times0.15''(\rm{10^\circ})$ & 2.45 \\
\dcn & 217.245846 & 1.34 & 20.9 & $0.18''\times0.14''(\rm{7^\circ})$ & 8.0 \\ 
\hline\hline
\end{tabular}
\end{table} 

\begin{figure}[h!]
\centering
    \includegraphics[width=\linewidth]{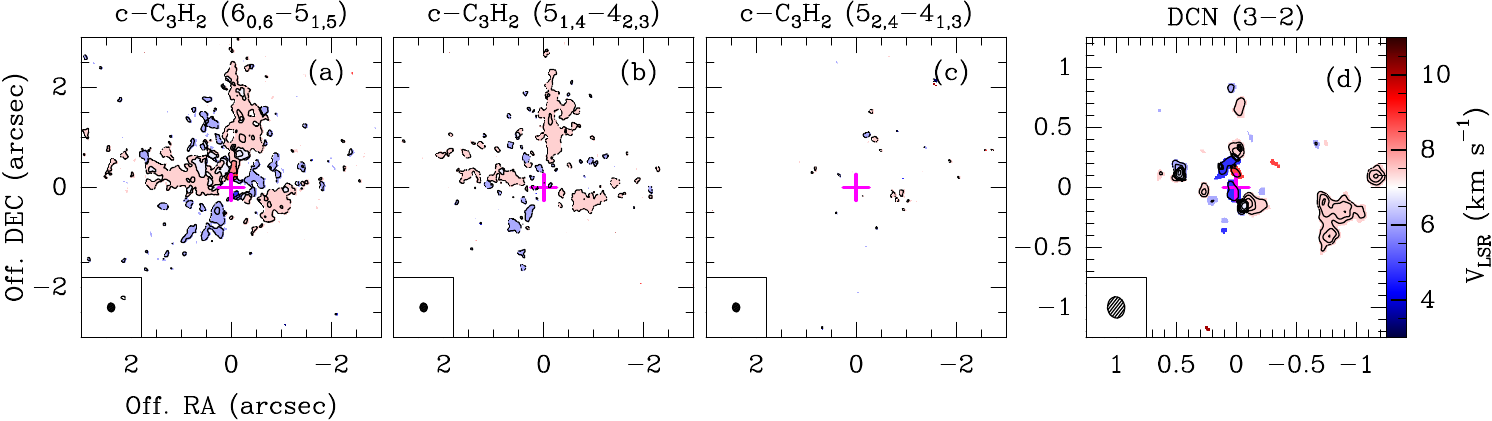}
    \caption{Moment 0 (contours) and moment 1 (color) maps of the c-C$_3$H$_2$ lines (a, b, c). In each panel, the green cross shows the position of the central protostar. The beam size of $0.17''\times0.14'', \rm{PA}=6^\circ$ is shown in the lower left corner and the contour levels are in steps of 3$\sigma$. \textit{Panel d:} Moment 0 (contours) and moment 1 (color) maps of the \dcn~ emission. The beam size of $0.18''\times0.14'', \rm{PA}=7^\circ$ is shown in the lower left corner. The contour levels start at $3\sigma$ and increase in steps of $6\sigma$.}\label{fig:c3h2+dcn}
\end{figure}

\begin{figure}[h!]
\centering
    \includegraphics[width=0.95\linewidth]{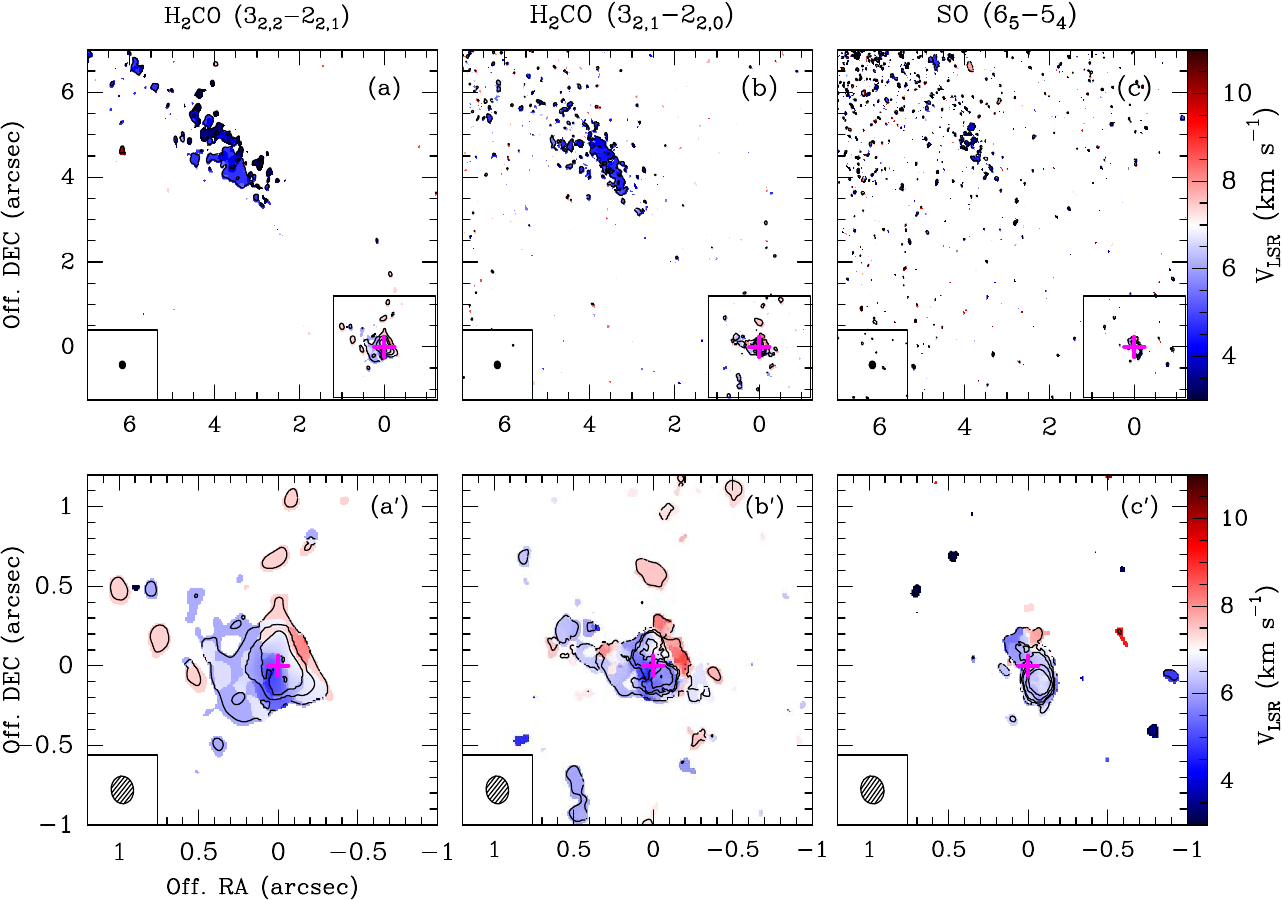}
    \caption{Moment 0 (contours) and moment 1 (color) maps of the \hhhhco~(a), \hhhco~(b), and \so~(c) emissions. The lower panel shows the same maps zoomed in to the central region marked by the black box in the upper panels. The beam size is shown in the lower left corner of each map: $0.17''\times0.14''(\rm{PA}=7^\circ)$ for the cases of \hhhco~(a, a$'$) and \hhhhco~(b, b$'$), and $0.18''\times0.15''(\rm{PA}=10^\circ)$ for the case of \so~(c, c$'$), respectively. The contour levels are $3\sigma, 9\sigma, 16\sigma, \rm{and}~ 25\sigma$.} \label{fig:h2co+so}
\end{figure}
\end{document}